# Comparative analysis of machine learning models for Ammonia Capture of Ionic Liquids

*Shahaboddin Shamshirband* [1,2], *Narjes Nabipour* [3*], *Masoud Hadipoor* [4], *Alireza Baghban* [5], *Amir Mosavi* [6,7,8,9*]

[1] *Department for Management of Science and Technology Development, Ton Duc Thang University, Ho Chi Minh City, Viet Nam*
[2] *Faculty of Information Technology, Ton Duc Thang University, Ho Chi Minh City, Viet Nam*
[3] *Institute of Research and Development, Duy Tan University, Da Nang 550000, Vietnam;*
[4] *Department of Petroleum Engineering, Ahwaz Faculty of Petroleum Engineering, Petroleum University of Technology (PUT), Ahwaz, Iran*
[5] *Department of Chemical Engineering, Amirkabir University of Technology, Mahshahr Campus, Mahshahr, Iran*
[6] *Kalman Kando Faculty of Electrical Engineering, Obuda University, Budapest 1034, Hungary*
[7] *School of the Built Environment, Oxford Brookes University, Oxford OX30BP, UK*
[8] *Institute of Structural Mechanics, Bauhaus Universität-Weimar, D−99423 Weimar, Germany*
[9] *Department of Mathematics and Informatics, J. Selye University, 94501 Komarno, Slovakia*

**Abstract –** Industry uses various solvents in the processes of refrigeration and ventilation. Among them, the Ionic liquids (ILs) as the relatively new solvents, are known for their proven eco-friendly characteristics. In this research, a comprehensive literature review was carried out to deliver an insight into the ILs and the prediction models used for estimating the ammonia solubility in ILs. Furthermore, a number of advanced machine learning methods, i.e. multilayer perceptron (MLP) and a combination of particle swarm optimization (PSO) and adaptive neuro-fuzzy inference system (ANFIS) models are used to estimate the solubility of ammonia in various ionic liquids. Affecting parameters were molecular weight, critical temperature and pressure of ILs. Furthermore, the salability is also predicted using two equation of states. Down the line, some comparisons were drawn between experimental and modelling results which is rarely done. It was seen that for MLP-ANN, PSO-ANFIS, Peng-Robinson and Soave-Redlich-Kwong equations of states, the R squares were 0.992, 0.9125, 0.8681, 0.4225 and Mean Square Errors were 0.00043, 0.004577, 0.010156, and 0.102345, respectively. In addition, study shows that the equations of

states (EOSs) couldn't estimate the solubility of ammonia accurately, by contrast artificial intelligence methods have produced promising results.

1. Introduction

The rapid economic growth and the ever increasing energy consumption, is posing a great challenge for the global sustainable development goals (Ardabili et al., 2018, Dehghani et al., 2019, Imani et al., 2018, Najafi et al., 2018, Nosratabadi et al., 2019). Thus, producing the energy from alternative energy resources of either sustainable or recycled materials have become very popular recently (Dineva et al., 2019, Moeini et al., 2018, Mosavi et al., 2019, Najafi et al., 2018, Torabi et al., 2019). One of the most noteworthy economical ways to recover and recycle low-quality waste heat in industrial and residential applications is low-grade energy utilization (Shi and Che, 2009). Low-quality waste heat was investigated and heat absorption in the refrigeration cycle was desired (Nitkiewicz and Sekret, 2014). Absorption chillers are important equipment to recapture waste heat. Refrigeration systems are adapted to use renewable energy sources (Abedin et al., 2014, Li et al., 2014, Cai et al., 2014b). Nowadays, however, the absorption refrigeration cycles are not very common because of increasing environmental issues and considerations (Bao et al., 2011, Junye et al., 2014).

An important class of non-aqueous solvents are ionic liquids (Cui and Schmidt, 2019, Larrechi et al., 2019, Zeng et al., 2019). Up to date, this kind of solvents has been widely used in industry and academic investigations (Kundu et al., 2019, Wang et al., 2019). ILs attracted attention due to their ability of reduction of organic solvents usage because some of the organic solvents are considered as pollutant materials. Ionic liquids have outstanding features which make them appropriate to be utilized in a wide range of new syntheses (Gogoi and Talukdar, 2014). The first ionic liquid was invented in the 1970s and used in some kind of batteries (Yilbas and Sahin, 2014) and now they are suggested as an alternative absorbent instead of previous working fluids (Cai et al., 2014a, Welton, 1999). Also, they have been widely used in industrial operations such as gas purification, heat transfer and elimination of metal ions (Gorman, 2001). Ionic liquids are able to dissolve many kinds of materials, their vapor pressures are extremely low and they are not miscible in the majority of organic solvents and often have a high level of thermal conductivity and thermal stability. (Li et al., 2010, Ayou et al., 2014, Brennecke and Maginn, 2001).

Up to date, huge number of researches were done to determine the solubility of gases when ILs exist in the solution (Yang and Dionysiou, 2004, Lagrost et al., 2003, Shariati and Peters, 2005, Yokozeki et al., 2008, Kurnia et al., 2009, Blanchard et al., 2001, Carvalho et al., 2009, Kim et al., 2011). The often researches provide data on the solubility of Carbon Dioxide and there are a few studies about the solubility of Ammonia (Wang et al., 2011, Hong et al., 2011, Jou and Mather, 2007). As is mentioned, just a few numbers of studies dealt with the topic of ammonia solubility in ionic liquids and this caused major problems for the authors to find and gather appropriate data bank. In spite of all difficulties, adequate number of data points have been found and the artificial intelligence system have been developed and was compared with the results of equations of state as the novelty of this work.

In addition to experimental studies, several studies and investigations were done to predict the gas solubility in ILs-bearing solutions. The most common approaches were neural networks and equation of states (Yokozeki and Shiflett, 2007a, Yokozeki and Shiflett, 2007b, Chen et al., 2013).

Two main subsets of neural networks are artificial neural networks (ANNs) and Fuzzy Logic System (FLS). ANNs are intelligence paradigms that are useful in the construction of strong and precise models. Several kinds of modifications of ANNs can be implemented to process a range of input data and predict output values with admirable precisions. ANNs are able to predict very complex and completely non-linear systems while they do not need any kind of simplifying assumptions (Haghbakhsh et al., 2013, Ashrafmansouri and Raeissi, 2012) for instance they can be easily utilized in absolutely complex problems of fluid flow, dissolution, evaporation, emission of pollutants and many other scientific areas. Yaseen et al in 2019 predicted the flow of river using an advanced method of artificial intelligence which can be easily integrated to the water engineering applications (Yaseen et al., 2018). In another outstanding paper, in 2018, Moazenzadeh et al have investigated o the application of SVR method to predict the evaporation rate in two distinct meteorological zones (Rasht and Lahijan, Iran)(Moazenzadeh et al., 2018). Chen et al in 2016 investigated the efficiency of a hybrid neural network in the estimation of sediment loads (Chen and Chau, 2016). The method of Quantum-based PSO was used in a paper by Ghorbani et al in 2018, which predicted the amount of pan evaporation in Talesh, Iran (Ali Ghorbani et al., 2018).

On the other hand, FLS are conceptually easy to understand because of very simple mathematical concepts and conditional statements are always TRUE or FALSE (Lashkarbolooki et al., 2013). By the combination of artificial neural networks (ANNs) (Cybenko, 1989, Baghban and Adelizadeh, 2018) and Fuzzy Logic (Zendehboudi et al., 2012), an adaptive neuro-fuzzy inference system which is named ANFIS can be developed. Back Propagation (BP) procedure is an optimization method which simultaneously influences the learning ability of model and leads to more accurate results. Some other optimization methods such as Genetic Algorithm, Non-dominated Sorting Genetic Algorithm (NSGA), Particle Swarm Optimization (PSO), Hybrid Particle Swarm Optimization and Genetic Algorithm (HPSOGA) and so forth can be used to do optimization (Baghban et al., 2018, Baghban et al., 2019b). In the current study, PSO-ANFIS, MLP-ANN, SRK and PR approaches were utilized to predict the solubility of Ammonia in eleven different solutions of ILs. Experimental conditions include many different values of temperatures, concentrations and pressures. Additionally, based on statistical approaches, a series of comparisons were made and modelling results were compared with experimental data.

## 2. Theory

### 2.1 Theory of ammonia solubility in ionic liquids

In recent decades, Ionic Liquids have been under considerable attention. These liquids have some properties which make them valuable and attractive for many scientific investigations. Due to their features, various types of them have been utilized in many researchers with a wide range of cations and anions. Furthermore, researchers have worked on the solubility of different gases such as ammonia, $CO_2$ and so forth and the effect on some parameters on their solubility. As long as their structure is concerned, ionic liquids are formed by a huge cation and an inorganic anion. As it is obvious, they are able to form many versatile structures and configurations by the combination of cations and anions. The process of obtaining the solubility and other properties of ionic liquids is expensive so many scientists have worked on correlations and artificial intelligence to find their features more rapidly and affordable.

At room temperatures, Ammonia can easily dissolve in ILs. In addition to that, by heating the sample or lowering the pressure, it can be desorbed in ILs (A Yokozeki, Shiflett, & research, 2007). Shi and Maginn (W. Shi and Maginn, 2009) have studied the absorption of ammonia in ILs between temperatures 298 and 348 K and interpreted the results which showed the high solubility ammonia in ILs. Results are highly conceivable and illustrate that NH3 forms relatively powerful hydrogen bonds between ring hydrogen atoms and nitrogen atoms. On the other hand, they stated that anion doesn't play an important role in solubility, which is contradictory with mechanisms of CO2 solubility in ILs. In a recent study, the effect of alkyl chain was investigated and it was found that by increasing the length of the alkyl, the solubility of ammonia will increase (G. Li, Zhou, Zhang, Zhang, & Li, 2010). Due to different types of ionic liquids, it is necessary to use a proper one to effectively dissolve ammonia. Regarding that, Palomar et al utilized (Bedia et al., 2012; Palomar et al., 2011) the COSMO-RS approach for finding the best IL for the absorption of ammonia. They used [choline]+[Tf2N]−, which has resulted in the best absorption performance.

The solubility of NH3 in ILs can be obtained using the concept of dissolved mole fraction of ammonia. Let suppose that mole of NH3 and IL is denoted by M1 and M2 then according to the following equations (when the number of components is N=2):

$$\overline{V_l} = 0.5 \sum_{i=1}^{N} (V_1^0 + V_2^0)(1 - m_{ij}) x_i x_j, \quad m_{ii} = 0, \quad m_{ij} = m_{ji} \qquad (1)$$

Where x is the mole fraction of each component, $\overline{V_l}$ is liquid molar volume, $V_T$, $V_i^0$ and mij are the total volume, individual saturated liquid molar volume of each component and the interaction parameter between components.

$$\overline{V_l} = V_1^0 x_1 + V_1^0 x_2 - m_{12}(V_1^0 + V_2^0) x_1 x_2 \qquad (2)$$

$$x_1 = \frac{M_{L1}}{M_{L1} + M_2}, \text{ and } x_2 = 1 - x_1 \qquad (3)$$

Where ML1 is the mole fraction of component one in liquid phase. Dg is the gaseous molar density (in units of mol/cm3)

$$V_L = (M_{L1} + M_2)\overline{V_l} \tag{4}$$

$$M_1 = D_s(V_T - V_l) + M_{L1} \tag{5}$$

$$aM_{L1}^2 + bM_{L1} + c = 0 \tag{6}$$

In above equations Ds is the amount of moles per cubic centimeter and is the gaseous molar density. Additionally, coefficients of a, b and c were defined as follows:

$$a = 1 - D_s V_1^0 \tag{7a}$$

$$b = D_s\left(V_T - M_2(V_1^0 + V_2^0)(1 - m_{12})\right) + M_2 - M_1 \tag{7b}$$

$$c = D_s M_2 (V_T - M_2 V_2^0) - M_1 M_2 \tag{7c}$$

By inserting predetermined factors of Ds, $V_i^0$ in equations and the results of equation 6 into equation 3, the solubility of ammonia in the specified IL will be determined (Yokozeki et al., 2007).

## 2.2 Fuzzy Logic System (FLS)

Field of FLS applications has been extended considerably. These systems have been used in various industries from programming of washing machines to petroleum industries (Yokozeki and Shiflett, 2007a, Yokozeki and Shiflett, 2007b). In the first place, Fuzzy logic theory was invented by Dr. Lotfi Zadeh in the 1960s. FLs systems and human relations closely resemble in some ways. Actually, Fuzzy logic (FL) is a logical system and is in close relationship to the fuzzy set theories, a theory which communicates to sets of input materials with not well-defined restrictions.

Due to absolutely simple mathematical concepts, Principles of FL are not very difficult to grasp. Also, FL is very versatile and can be easily matched with different systems. So it is capable of modelling hard and complex systems which cannot be analyzed with ordinary mathematical approaches (Chen et al., 2013).

In additional details, a FIS uses IF–Then conditional statements. They utilize a set of explicit rules which are used to convert implicit inputs in order to build a set of conditions determining the outputs (Brown and Gabbar, 2014). Meanwhile, Membership Functions must provide a transition path form real data to the heart of Fuzzy logic environment. Based on so many different problems with varying natures, different types of memberships can be used. The corresponding parameters of MFs can be adjusted numerically (Veit, 2012, Castillo and Melin, 2014). In Equations (1), (2) and (3), three types of most common MFs have been formulated.

Thus, designed fuzzy conditions were set to control systems with communicating between input and output variables(Rostami et al., 2019). Then, different modelling efforts were done to boost the accuracy and straightforwardness of the designed procedures(Baghban and Adelizadeh, 2018).

$$zmf(x;m,n) = \begin{bmatrix} 0, x \leq m \\ 2\left(\dfrac{x-m}{n-m}\right)^2, m \leq x \leq \dfrac{m+n}{2} \\ 1-2\left(\dfrac{x-n}{n-m}\right)^2, \dfrac{n+m}{2} \leq x \leq n \\ 0, x \geq n \end{bmatrix} \quad : zmf\ membership\ function \quad (8)$$

$$pimf(x;m,n,i,k) = \begin{bmatrix} 0, x \leq m \\ 2\left(\dfrac{x-m}{n-m}\right)^2, m \leq x \leq \dfrac{m+n}{2} \\ 1-2\left(\dfrac{x-n}{n-m}\right)^2, \dfrac{n+m}{2} \leq x \leq n \\ 1, n \leq x \leq i \\ 1-2\left(\dfrac{x-i}{k-i}\right)^2, i \leq x \leq \dfrac{i+k}{2} \\ 2\left(\dfrac{x-k}{k-i}\right)^2, \dfrac{k+i}{2} \leq x \leq k \\ 0, x \geq k \end{bmatrix} \quad \text{: pimf membership function} \quad (9)$$

$$smf(x;m,n) = \begin{bmatrix} 1, x \leq m \\ 1-2\left(\dfrac{x-m}{n-m}\right)^2, m \leq x \leq \dfrac{m+n}{2} \\ 2\left(\dfrac{x-m}{n-m}\right)^2, \dfrac{m+n}{2} \leq x \leq n \\ 1, x \geq n \end{bmatrix} \quad \text{: smf membership function} \quad (10)$$

## 2.3 Artificial Neural Networks

Ahmadi (Ahmadi et al., 2014) states that a neural network (NN) has a tendency to use processing units in a straightforward manner to process inputted data and store experimental information to be available for later use. The most used and prevalent model of Artificial Neural Network in many practical applications is a Feed Forward neural network which is the earlier version of ANNs (Anifowose and Abdulraheem, 2011, Helmy et al., 2010, Haykin, 1994). They use neither cycles nor loops and the information just go forward. Single-layer perceptron networks are the simplest kind of neural networks. Data directly enter the output layer and the only things that influence the data are weighting parameters (Svozil et al., 1997).

The multi-layer perceptron (MLP) networks contain different layers. An activation function, named sigmoid function, is implemented in numerous modelling efforts. Transfer functions are presented in Equations (4), (5) and (6) where equation (5) is the Linear (purelin) one, equation (6) is showing the Log-Sigmoid (logsig) one, and equation (6) represent Hyperbolic Tangent Sigmoid one (tansig).

$$\phi(k) = k \tag{11}$$

$$\phi(m) = \frac{1}{1 + e^{-k}} \tag{12}$$

$$\phi(m) = \frac{e^k - e^{-k}}{e^k + e^{-k}} \tag{13}$$

In addition to input and output layers, the multi-layer perceptron usually has a set of interior layers which are hidden and direct access to them is not possible (Samavati et al., 2013, Sipöcz et al., 2011, Baghban et al., 2019a). A trial and error approach must be used to obtain the best figures of interior layers. However, the training part of the model will not operate properly if the figures of interior neurons are more than the total neurons in the network. This phenomenon greatly decreases the accuracy of the results. On the other side, if the neurons in hidden layers are less than those in the network, the process will remain immature because of rapid and unexpected convergence.

## 2.4  ANFIS

By merging ANNs with fuzzy logic systems, ANFIS system is created (Fan et al., 2018a, Fan et al., 2018b, Kaur and Sood, 2019, Polykretis et al., 2019). ANFIS is formed based on a well-known subset of ANN systems, i.e., back-propagation and a fuzzy optimization technique encounter training the parameters of the FIS membership functions (Abdillah and Suharjito, 2019, Kumar, 2019, Rezakazemi and Shirazian, 2019). Based on a given error criterion, the system parameters changed over and over until they met the criteria. Afterwards, the ANFIS system, or so called the adaptive neuro fuzzy inference system, learns by extracting the main features of the input data (Riahi-Madvar, et al. 2019; Mosavi et al. 2019). Error criterion defined to be an array of RMS errors which signify the error signal of training data (Mirarab et al., 2014). ANFIS has gained popularity in a wide range of applications (Ardabili et al., 2018, Mosavi and Edalatifar, 2019, Mosavi et al., 2018, Mosavi et al., 2019, Rezakazemi et al., 2019) .

## 2.5    Particle Swarm Optimization

The PSO algorithm is an optimization approach which at the first place, was developed in 1995 by Kennedy and Eberhart (Arriagada et al., 2002). Movement of a bunch of birds forms the principal concept of PSO when they are flocking somewhere. This algorithm creates a random crowd and introduces modification and changes to this crowd until reaches an optimal situation. There are no evolution operators in PSO method while Genetic algorithms use them. In PSO method particles form the crowd and every particle travels along with adjustable speeds and direction. For every molecule, the speed and direction are changed over and over. Then, previously mentioned velocity is supplemented to the situation of the molecule. Speed changes are exposed to both the best worldwide hopeful path and the best close to the path of every molecule. If the optimized path and cost of the particle is more effective than the cost of the global path, then the global path will be substituted by the individual particle path (Sadrzadeh et al., 2009).

## 2.6    PSO-ANFIS

Due to the accuracy and computational efficiency of PSO-ANFIS it has recently gained popularity among scientific communities and various application domains (Abbasi and Hadji Hosseinlou, 2019, Ceylan et al., 2018, Darvish et al., 2018, Keybondorian et al., 2018, Liu et al., 2018, Malmir et al., 2018, Mir et al., 2018, Mottahedi et al., 2018, Rezakazemi et al., 2017, Suleymani and Bemani, 2018). Basser et al. (2015) and Shamshirband et al. (2019) provide a comprehensive description of PSO-ANFIS.

## 2.7 Equations of state

Equations of states (EOS) are thermodynamic equations to express the state of substances in different situations. This equation has widely used in industries (where properties of fluids are the matter of interest), physics (to determine the state of the interior of stars) and other fields. Redlich–Kwong, Peng–Robinson, Soave Redlich–Kwong and Patel–Teja Valderrama and Patel–Teja are some examples of EOSs (Baghban et al., 2015).

### 2.7.1 Peng-Robinson EOS

A very useful and important EOS was proposed by Peng and Robinson (Kennedy and Eberhart, 1995) which is mentioned in eq. (7)

$$P = \frac{RT}{V-b} + \frac{a(T)}{V(V-b)+b(V-b)} \tag{14}$$

$$b_i = \frac{0.077796 \, RT_{ci}}{P_{ci}} \tag{15}$$

The mixture parameters can be calculated as follows:

$$a = \sum_i \sum_j x_i x_j a_{ij} \tag{16}$$

$$a_{ij} = (a_{ii} a_{jj})^{0.5}(1-k_{ij}) \tag{17}$$

$$b = \sum_i \sum_j x_i x_j b_{ij} \tag{18}$$

$$b_{ij} = (\frac{b_i + b_j}{2})(1-l_{ij}) \tag{19}$$

$l_{ij}$ and $k_{ij}$ represent the interaction factors in Eqs. (10) and (12) and it is noteworthy to mention that in Eq. (11), $b_{ii} = b_i$ and $b_{jj} = b_j$.

$$a_{ii} = \frac{0.457235 R^2 T_{ci}^2}{P_{ci}} \left[1+(0.37464+1.54226\omega_i - 0.26992\omega^2)\left(1-\sqrt{\frac{T}{T_{ci}}}\right)\right]^2 \tag{20}$$

$$b_i = \frac{0.077796 RT_{ci}}{P_{ci}} \tag{21}$$

Where $T_c$ stands for critical temperature, $P_c$ represents critical pressure and $\omega$ is the acentric factor.

### 2.7.2 Soave-Redlich-Kwong

One another powerful EOSs was presented by Soave et al(Kennedy and Eberhart, 1995):

$$P = \frac{RT}{V-b} + \frac{a(T)}{V(V+b)} \tag{22}$$

Where its parameters a and b were proposed as follows:

$$b = \frac{0.08664\,RT_c}{P_c} \tag{23}$$

$$a = \frac{0.4274R^2T_c^2}{P_c}\left[1+m\left(1-\sqrt{\frac{T}{T_c}}\right)\right]^2 \tag{24}$$

$$m = 0.480 + 1.57\omega - 0.176\omega^2 \tag{25}$$

## 3. Data preparation and analysis

To develop and train two previously mentioned neural networks, a set of precise experimental data is needed. In this regard, a survey was done to check previously published papers in order to gather Ammonia solubility data. Eventually, 352 experimental data were obtained. Experimental details such as mole fractions, ionic liquids, pressures and so on are represented in Table 1. A range of detailed thermodynamic properties which is used in the present work is mentioned in Table 2. Approximately 75% of obtained data were used for training purpose and about 25% of them were used for testing procedures. Furthermore, the efficiencies of the proposed models were evaluated by using testing data group.

## 4. Executions of networks training

In present work, the PSO-ANFIS and the MLP-ANN models were merged with each other to estimate the solubility of Ammonia in the presence of different kinds of ionic liquids. Diagram of proposed ANN and ANFIS models, which indicates how the models were developed, was presented in **Fig. 1** for ammonia solubility in ILs.

In order to combine PSO-ANFIS and MLP-ANN models, a programming attempt was done in MATLAB software. The inputted and desired data values were normalized and the training part was done by adding a Bayesian technique to Levenberg–Marquardt algorithm (Prausnitz et al., 1998). MLP-ANN was implemented using a Log-Sigmoid transfer function in the invisible layers and linear transfer function in the output layer. An iterative procedure was utilized to determine the optimal value for the number of interior layers. PSO-ANFIS was run based on Gaussian membership function and with the end goal of optimization of FIS parameters, particle swarm optimization method (PSO) was employed.

An evaluation of the Performance of each model was done based on using statistical analysis. Some of these technics were MSEs, AAEs, STD, RMSEs, R-Squared and also, the histogram of the errors. Equations (19) to (23) demonstrate the formulation of statistical approaches.

$$MSE = \frac{1}{k}\sum_{i=1}^{k}(a_{\exp} - a_{cal})^2 \qquad (26)$$

$$AAE = \frac{1}{k}\sum_{i=1}^{k}|a_{\exp} - a_{cal}| \qquad (27)$$

$$STD = \left(\frac{1}{k-1}\sum_{i=1}^{k}(\varepsilon - \bar{\varepsilon})^2\right)^{0.5} \qquad (28)$$

$$RMSE = \left(\frac{1}{k}\sum_{i=1}^{k}(a_{\exp} - a_{cal})^2\right)^{0.5} \qquad (29)$$

$$R^2 = 1 - \frac{\sum_{i=1}^{k}(a_{\exp} - a_{cal})^2}{\sum_{i=1}^{k}(a_{\exp} - \bar{a}_{\exp})^2} \qquad (30)$$

Where $a_{\exp}$, $a_{cal}$, k, $\varepsilon$ are experimental data, estimated values, total number of data and errors, respectively.

## 5. Results and discussion

Present work uses PSO-ANFIS, MLP-ANN, PR-EOS and SRK-EOS to predict and conceptualize the relationship among input and desired outputs. From 352 experimental data presented in the literature about the solubility of Ammonia, 264 data points were utilized to train the models and 88 data were considered as testing data. Test data set was not entered the training procedure. It is noteworthy to mention that the training data were also employed in the testing section. Moreover, experimentally obtained data were completely employed for prediction of ammonia solubility by SRK and PR models. With the end goal of finding answers using EOSs, acentric factors, critical properties of ionic liquids and fugacity data of both phases were used to determine a and b coefficients in the utilized equations of state. It must be noted that in these calculations, for simplicity, we ignored the interaction terms between NH3 and ILs and calculated the results of EOSs. Development of the PSO-ANFIS model was based on Gaussian membership function. A ten clusters configuration was used and 120 FIS parameters were achieved. Optimization was done according to the PSO method to optimize the outputs of PSO-ANFIS model. Additionally, the configuration of the MLP network was done according to the best number of interior layers and those of neurons in each layer. To determine how many interior layer neurons should be used, an iterative method was considered. The structure of MLP-ANN has been illustrated in Fig.2. Specifications of training procedures for PSO-ANFIS and MLP-ANN systems are given in Tables 3 and 4. Table 5 represents the weighting values and bias numbers of MLP-ANN network.

Minimization of the sum of the squared errors was continued for PSO-ANFIS and MLP-ANN until a steady response reached and no further changes in values occurred during iterative steps. To determine the efficiency of proposed models, the MSEs, AAEs, STDs, RMSEs, R-Squared (R2) and also, the histogram of the errors were used.

Experimental data and model results were evaluated for EOSs, PSO-ANFIS and MLP–ANN systems as demonstrated in Figs. 3- 5. According to the obtained model results, MLP-ANN had outstanding precision in predicting the Ammonia solubility. Resulted values from PSO-ANFIS and EOS systems were not as precise as those from MLP-ANN method. Figs. 6-8 illustrate the regression of the results of each model versus experimental values. As is mentioned before, MLP-ANN results were approximately fitted to the experimental results and a great agreement was seen. The values of R2 for MLP-ANN and PSO-ANFIS models were 0.9967, 0.992 and 0.9618, 0.9125 for training and testing stages of models, respectively. Additional, values of 0.8681 and 0.4225 were obtained for R2 values of PR-EOS and SRK-EOS, respectively.

According to the employed statistical methods for EOSs, Equation (24) was obtained for PR-EOS and equation (25) was found for SRK-EOS. Down the line, regression of MLP-ANN was found to be according to equation (26) and the regression of PSO-ANFIS was according to equation (27).

$$y = 0.8735 x + 0.1048; R^2 = 0.8681 \tag{31}$$

$$y = 0.677x + 0.3499; R^2 = 0.4225 \tag{32}$$

$$y = 0.9994x + 0.0003; R^2 = 0.992 \tag{33}$$

$$y = 0.9653x + 0.0186; R^2 = 0.9125 \tag{34}$$

In Figs. 9-10 histograms of training and testing data set errors for both PSO-ANFIS and MLP-ANN models is illustrated. Also, in Fig. 11, histograms of errors for PR-EOS and SRK-EOS is presented.

Results of absolute deviation analysis of experimental values and results of presented models are demonstrated in Fig.12. As can be seen, the absolute deviation of artificial intelligence method is dramatically lower than those of PR and SRK equations of state. In addition to that, the accuracy of EOSs is highly dependent on the mole fraction of NH3 in ILs. By contrast, the results of artificial intelligence approaches remain precise. The percentage of absolute average errors of PSO-ANFIS model were 3.186571 and 1.629359. Corresponding values for MLP-ANN model were 1.23697 and 0.742338 for testing and training sets, respectively. While the percentage of absolute average errors of PR-EOS was 5.63, the similar parameter was obtained equal to 22.49 for SRK-EOS. Some additional statistical details of employed methods are presented in Table 6.

## 6. Conclusion

Experimental data from the literature were obtained. A feedforward Multi-Layer Perceptron (MLP) neural network and a combination of PSO optimization method and ANFIS system were proposed to develop a PSO-ANFIS. Ammonia solubility in 13 ionic liquids (ILs) in different conditions were modelled. Input variables were the molecular weight (Mw) of ILs, Pc $T_c$, pressure and temperature data. The concentration of Ammonia in solution was the desired output. Some outstanding EOSs such as PR and SRK models were used to find the solubility of Ammonia in ILs. The advantages and characteristics of the previously mentioned computational models result in a precise and fast approach to the final results. in addition to the quickness of models, obtaining final results is absolutely cheaper than the other methods. According to the outputs of developed models, it was found that MLP-ANN approach had a great superiority over PSO-ANFIS method and EOSs. Although a big data bank on the solution of other gases in ILs is available in literature, but shortage of appropriate data on the solubility of ammonia was the most important limitation in this study and for the future works it is highly recommended that the researcher should expand the data bank and try other types of equations of state and examine the solubility of other gases which are not studied well until now.

**Acknowledgement**

We acknowledge the financial support of this work by the Hungarian State and the European Union under the EFOP-3.6.1-16-2016-00010 project and the 2017-1.3.1-VKE-2017-00025 project.

**References**

Abbasi, E. & Hadji Hosseinlou, M. 2019. The Importance of Exercise and General Mental Health on Prediction of Property-Damage-Only Accidents among Taxi Drivers in Tehran: A Study Using ANFIS-PSO and Regression Models. *Journal of Advanced Transportation,* 2019.
Abdillah, Y. & Suharjito 2019. Failure prediction of e-banking application system using Adaptive Neuro Fuzzy Inference System (ANFIS). *International Journal of Electrical and Computer Engineering,* 9**,** 667-675.
Abedin, M. J., Masjuki, H. H., Kalam, M. A., Sanjid, A. & Ashraful, A. M. 2014. Combustion, performance, and emission characteristics of low heat rejection engine operating on various biodiesels and vegetable oils. *Energy Conversion and Management,* 85**,** 173-189.
Ahmadi, M. A., Ebadi, M., Marghmaleki, P. S. & Fouladi, M. M. 2014. Evolving predictive model to determine condensate-to-gas ratio in retrograded condensate gas reservoirs. *Fuel,* 124**,** 241-257.
Ali Ghorbani, M., Kazempour, R., Chau, K.-W., Shamshirband, S. & Taherei Ghazvinei, P. J. E. A. o. C. F. M. 2018. Forecasting pan evaporation with an integrated artificial neural network quantum-behaved particle swarm optimization model: A case study in Talesh, Northern Iran. 12**,** 724-737.
Anifowose, F. & Abdulraheem, A. 2011. Fuzzy logic-driven and SVM-driven hybrid computational intelligence models applied to oil and gas reservoir characterization. *Journal of Natural Gas Science and Engineering,* 3**,** 505-517.
Ardabili, S. F., Najafi, B., Alizamir, M., Mosavi, A., Shamshirband, S. & Rabczuk, T. 2018. Using SVM-RSM and ELM-RSM approaches for optimizing the production process of methyl and ethyl esters. *Energies,* 11.
Arriagada, J., Olausson, P. & Selimovic, A. 2002. Artificial neural network simulator for SOFC performance prediction. *Journal of Power Sources,* 112**,** 54-60.
Ashrafmansouri, S.-S. & Raeissi, S. 2012. Modeling gas solubility in ionic liquids with the SAFT-γ group contribution method. *The Journal of Supercritical Fluids,* 63**,** 81-91.
Ayou, D. S., Currás, M. R., Salavera, D., García, J., Bruno, J. C. & Coronas, A. 2014. Performance analysis of absorption heat transformer cycles using ionic liquids based on imidazolium cation as absorbents with 2,2,2-trifluoroethanol as refrigerant. *Energy Conversion and Management,* 84**,** 512-523.

Baghban, A. & Adelizadeh, M. J. F. 2018. On the determination of cetane number of hydrocarbons and oxygenates using Adaptive Neuro Fuzzy Inference System optimized with evolutionary algorithms. 230**,** 344-354.

Baghban, A., Ahmadi, M. A. & Hashemi Shahraki, B. 2015. Prediction carbon dioxide solubility in presence of various ionic liquids using computational intelligence approaches. *The Journal of Supercritical Fluids,* 98**,** 50-64.

Baghban, A., Jalali, A., Shafiee, M., Ahmadi, M. H. & Chau, K.-w. J. E. A. o. C. F. M. 2019a. Developing an ANFIS-based swarm concept model for estimating the relative viscosity of nanofluids. 13**,** 26-39.

Baghban, A., Kahani, M., Nazari, M. A., Ahmadi, M. H., Yan, W.-M. J. I. J. o. H. & Transfer, M. 2019b. Sensitivity analysis and application of machine learning methods to predict the heat transfer performance of CNT/water nanofluid flows through coils. 128**,** 825-835.

Baghban, A., Kardani, M. N. & Mohammadi, A. H. J. F. 2018. Improved estimation of Cetane number of fatty acid methyl esters (FAMEs) based biodiesels using TLBO-NN and PSO-NN models. 232**,** 620-631.

Bao, H. S., Wang, R. Z. & Wang, L. W. 2011. A resorption refrigerator driven by low grade thermal energy. *Energy Conversion and Management,* 52**,** 2339-2344.

Basser, H., Karami, H., Shamshirband, S., Akib, S., Amirmojahedi, M., Ahmad, R., Jahangirzadeh, A. & Javidnia, H. 2015. Hybrid ANFIS-PSO approach for predicting optimum parameters of a protective spur dike. *Applied Soft Computing Journal,* 30**,** 642-649.

Blanchard, L. A., Gu, Z. & Brennecke, J. F. 2001. High-Pressure Phase Behavior of Ionic Liquid/CO2 Systems. *The Journal of Physical Chemistry B,* 105**,** 2437-2444.

Brennecke, J. F. & Maginn, E. J. 2001. Ionic liquids: Innovative fluids for chemical processing. *AIChE Journal,* 47**,** 2384-2389.

Brown, C. & Gabbar, H. A. 2014. Fuzzy logic control for improved pressurizer systems in nuclear power plants. *Annals of Nuclear Energy,* 72**,** 461-466.

Cai, D., He, G., Tian, Q. & Tang, W. 2014a. Exergy analysis of a novel air-cooled non-adiabatic absorption refrigeration cycle with NH3–NaSCN and NH3–LiNO3 refrigerant solutions. *Energy Conversion and Management,* 88**,** 66-78.

Cai, D., He, G., Tian, Q. & Tang, W. 2014b. Thermodynamic analysis of a novel air-cooled non-adiabatic absorption refrigeration cycle driven by low grade energy. *Energy Conversion and Management,* 86**,** 537-547.

Carvalho, P. J., Álvarez, V. H., Schröder, B., Gil, A. M., Marrucho, I. M., Aznar, M., Santos, L. M. N. B. F. & Coutinho, J. A. P. 2009. Specific Solvation Interactions of CO2 on Acetate and Trifluoroacetate Imidazolium Based Ionic Liquids at High Pressures. *The Journal of Physical Chemistry B,* 113**,** 6803-6812.

Castillo, O. & Melin, P. 2014. A review on interval type-2 fuzzy logic applications in intelligent control. *Information Sciences,* 279**,** 615-631.

Ceylan, Z., Pekel, E., Ceylan, S. & Bulkan, S. 2018. Biomass higher heating value prediction analysis by ANFIS, PSO-ANFIS and GA-ANFIS. *Global Nest Journal,* 20**,** 589-597.

Chen, W., Liang, S., Guo, Y., Gui, X. & Tang, D. 2013. Investigation on vapor–liquid equilibria for binary systems of metal ion-containing ionic liquid [bmim]Zn2Cl5/NH3 by experiment and modified UNIFAC model. *Fluid Phase Equilibria,* 360**,** 1-6.

Chen, X. Y. & Chau, K. W. J. W. r. m. 2016. A hybrid double feedforward neural network for suspended sediment load estimation. 30**,** 2179-2194.

Cui, K. & Schmidt, J. R. 2019. Comment on "Solubilities of ammonia in basic imidazolium ionic liquids" [Fluid Phase Equilib. 297 (2010) 34–39]. *Fluid Phase Equilibria,* 492**,** 78-79.

Cybenko, G. 1989. Approximation by superpositions of a sigmoidal function. *Mathematics of Control, Signals and Systems,* 2**,** 303-314.

Darvish, H., Rahmani, S., Maleki Sadeghi, A. & Emami Baghdadi, M. H. 2018. The ANFIS-PSO strategy as a novel method to predict interfacial tension of hydrocarbons and brine. *Petroleum Science and Technology,* 36**,** 654-659.

Dehghani, M., Riahi-Madvar, H., Hooshyaripor, F., Mosavi, A., Shamshirband, S., Zavadskas, E. K. & Chau, K. W. 2019. Prediction of hydropower generation using grey Wolf optimization adaptive neuro-fuzzy inference system. *Energies,* 12.

Dineva, A., Mosavi, A., Ardabili, S., Vajda, I., Shamshirband, S., Rabczuk, T. & Chau, K. W. 2019. Review of soft computing models in design and control of rotating electrical machines. *Energies,* 12.

Fan, B., Lin, C., Wang, F., Liu, S., Liu, L. & Xu, S. 2018a. An Adaptive Neuro-Fuzzy Inference System (ANFIS) Based Model for the Temperature Prediction of Lithium-Ion Power Batteries. *SAE International Journal of Passenger Cars - Electronic and Electrical Systems,* 12.

Fan, B., Lin, C., Wang, F., Liu, S., Liu, L. & Xu, S. 2018b. An Adaptive Neuro-Fuzzy Inference System (ANFIS) Based Model for the Temperature Prediction of Lithium-Ion Power Batteries. *SAE International Journal of Passenger Cars - Electronic and Electrical Systems,* 11.

Gogoi, T. K. & Talukdar, K. 2014. Exergy based parametric analysis of a combined reheat regenerative thermal power plant and water–LiBr vapor absorption refrigeration system. *Energy Conversion and Management,* 83**,** 119-132.

Gorman, J. 2001. Faster, better, cleaner?: New liquids take aim at old-fashioned chemistry. *Science News,* 160**,** 156-158.

Haghbakhsh, R., Soleymani, H. & Raeissi, S. 2013. A simple correlation to predict high pressure solubility of carbon dioxide in 27 commonly used ionic liquids. *The Journal of Supercritical Fluids,* 77**,** 158-166.

Haykin, S. 1994. *Neural Networks: A Comprehensive Foundation*.

Helmy, T., Fatai, A. & Faisal, K. 2010. Hybrid computational models for the characterization of oil and gas reservoirs. *Expert Systems with Applications,* 37**,** 5353-5363.

Hong, S. Y., Im, J., Palgunadi, J., Lee, S. D., Lee, J. S., Kim, H. S., Cheong, M. & Jung, K.-D. 2011. Ether-functionalized ionic liquids as highly efficient SO2 absorbents. *Energy & Environmental Science,* 4**,** 1802-1806.

Imani, M. H., Zalzar, S., Mosavi, A. & Shamshirband, S. 2018. Strategic Behavior of Retailers for Risk Reduction and Profit Increment via Distributed Generators and Demand Response Programs. *Energies,* 11.

Jou, F.-Y. & Mather, A. 2007. Solubility of Hydrogen Sulfide in [bmim][PF6]. *International Journal of Thermophysics,* 28**,** 490-495.

Junye, H., Yaping, C. & Jiafeng, W. 2014. Thermal performance of a modified ammonia–water power cycle for reclaiming mid/low-grade waste heat. *Energy Conversion and Management,* 85**,** 453-459.

Kaur, H. & Sood, S. K. 2019. Adaptive Neuro Fuzzy Inference System (ANFIS) based wildfire risk assessment. *Journal of Experimental and Theoretical Artificial Intelligence,* 31**,** 599-619.

Kennedy, J. & Eberhart, R. Particle swarm optimization.  Neural Networks, 1995. Proceedings., IEEE International Conference on, Nov/Dec 1995 1995. 1942-1948 vol.4.

Keybondorian, E., Taherpour, A., Bemani, A. & Hamule, T. 2018. Application of novel ANFIS-PSO approach to predict asphaltene precipitation. *Petroleum Science and Technology,* 36**,** 154-159.

Kim, J. E., Lim, J. S. & Kang, J. W. 2011. Measurement and correlation of solubility of carbon dioxide in 1-alkyl-3-methylimidazolium hexafluorophosphate ionic liquids. *Fluid Phase Equilibria,* 306**,** 251-255.

Kumar, S. 2019. Estimation capabilities of biodiesel production from algae oil blend using adaptive neuro-fuzzy inference system (ANFIS). *Energy Sources, Part A: Recovery, Utilization and Environmental Effects*.

Kundu, D., Chakma, S., Pugazhenthi, G. & Banerjee, T. 2019. Reactive insights into the hydrogen production from ammonia borane facilitated by phosphonium based ionic liquid. *Korean Journal of Chemical Engineering,* 36**,** 456-467.

Kurnia, K. A., Harris, F., Wilfred, C. D., Abdul Mutalib, M. I. & Murugesan, T. 2009. Thermodynamic properties of $CO_2$ absorption in hydroxyl ammonium ionic liquids at pressures of (100–1600) kPa. *The Journal of Chemical Thermodynamics,* 41**,** 1069-1073.

Lagrost, C., Carrié, D., Vaultier, M. & Hapiot, P. 2003. Reactivities of Some Electrogenerated Organic Cation Radicals in Room-Temperature Ionic Liquids:  Toward an Alternative to Volatile Organic Solvents? *The Journal of Physical Chemistry A,* 107**,** 745-752.

Larrechi, M. S., Cera-Manjarres, A. & Coronas, A. 2019. Ranking the solubility of ammonia in ionic liquids using near infrared spectroscopy and multivariate curve resolution. *Spectrochimica Acta - Part A: Molecular and Biomolecular Spectroscopy,* 215**,** 88-96.

Lashkarbolooki, M., Shafipour, Z. S., Hezave, A. Z. & Farmani, H. 2013. Use of artificial neural networks for prediction of phase equilibria in the binary system containing carbon dioxide. *The Journal of Supercritical Fluids,* 75**,** 144-151.

Li, G., Zhou, Q., Zhang, X., LeiWang, Zhang, S. & Li, J. 2010. Solubilities of ammonia in basic imidazolium ionic liquids. *Fluid Phase Equilibria,* 297**,** 34-39.

Li, X., Huang, H. & Zhao, W. 2014. A supercritical or transcritical Rankine cycle with ejector using low-grade heat. *Energy Conversion and Management,* 78**,** 551-558.

Liu, Z., Zhang, D. & Peng, W. 2018. A Novel ANFIS-PSO Network for forecasting oil flocculated asphaltene weight percentage at wide range of operation conditions. *Petroleum Science and Technology,* 36**,** 1044-1050.

Malmir, P., Suleymani, M. & Bemani, A. 2018. Application of ANFIS-PSO as a novel method to estimate effect of inhibitors on Asphaltene precipitation. *Petroleum Science and Technology,* 36**,** 597-603.

Marquardt, D. W. 1963. An Algorithm for Least-Squares Estimation of Nonlinear Parameters. *Journal of the Society for Industrial and Applied Mathematics,* 11**,** 431-441.

Mir, M., Kamyab, M., Lariche, M. J., Bemani, A. & Baghban, A. 2018. Applying ANFIS-PSO algorithm as a novel accurate approach for prediction of gas density. *Petroleum Science and Technology,* 36**,** 820-826.

Mirarab, M., Sharifi, M., Ghayyem, M. A. & Mirarab, F. 2014. Prediction of solubility of $CO_2$ in ethanol–[EMIM][Tf2N] ionic liquid mixtures using artificial neural networks based on genetic algorithm. *Fluid Phase Equilibria,* 371**,** 6-14.

Moazenzadeh, R., Mohammadi, B., Shamshirband, S. & Chau, K.-w. J. E. A. o. C. F. M. 2018. Coupling a firefly algorithm with support vector regression to predict evaporation in northern Iran. 12**,** 584-597.

Moeini, I., Ahmadpour, M., Mosavi, A., Alharbi, N. & Gorji, N. E. 2018. Modeling the time-dependent characteristics of perovskite solar cells. *Solar Energy,* 170**,** 969-973.

Mosavi, A. & Edalatifar, M. 2019. A Hybrid Neuro-Fuzzy Algorithm for Prediction of Reference Evapotranspiration. *Lecture Notes in Networks and Systems.* Springer.

Mosavi, A., Ozturk, P. & Chau, K. W. 2018. Flood prediction using machine learning models: Literature review. *Water (Switzerland),* 10.

Mosavi, A., Salimi, M., Ardabili, S. F., Rabczuk, T., Shamshirband, S. & Varkonyi-Koczy, A. R. 2019. State of the art of machine learning models in energy systems, a systematic review. *Energies,* 12.

Mottahedi, A., Sereshki, F. & Ataei, M. 2018. Overbreak prediction in underground excavations using hybrid ANFIS-PSO model. *Tunnelling and Underground Space Technology,* 80**,** 1-9.

Najafi, B., Ardabili, S. F., Mosavi, A., Shamshirband, S. & Rabczuk, T. 2018. An intelligent artificial neural network-response surface methodology method for accessing the optimum biodiesel and diesel fuel blending conditions in a diesel engine from the viewpoint of exergy and energy analysis. *Energies,* 11.

Nitkiewicz, A. & Sekret, R. 2014. Comparison of LCA results of low temperature heat plant using electric heat pump, absorption heat pump and gas-fired boiler. *Energy Conversion and Management,* 87**,** 647-652.

Nosratabadi, S., Mosavi, A., Shamshirband, S., Zavadskas, E. K., Rakotonirainy, A. & Chau, K. W. 2019. Sustainable business models: A review. *Sustainability (Switzerland),* 11.

Peng, D.-Y. & Robinson, D. B. 1976. A New Two-Constant Equation of State. *Industrial & Engineering Chemistry Fundamentals,* 15**,** 59-64.

Polykretis, C., Chalkias, C. & Ferentinou, M. 2019. Adaptive neuro-fuzzy inference system (ANFIS) modeling for landslide susceptibility assessment in a Mediterranean hilly area. *Bulletin of Engineering Geology and the Environment,* 78**,** 1173-1187.

Prausnitz, J. M., Lichtenthaler, R. N. & de Azevedo, E. G. 1998. *Molecular Thermodynamics of Fluid-Phase Equilibria*, Pearson Education.

Rezakazemi, M., Dashti, A., Asghari, M. & Shirazian, S. 2017. H2-selective mixed matrix membranes modeling using ANFIS, PSO-ANFIS, GA-ANFIS. *International Journal of Hydrogen Energy,* 42**,** 15211-15225.

Rezakazemi, M., Mosavi, A. & Shirazian, S. 2019. ANFIS pattern for molecular membranes separation optimization. *Journal of Molecular Liquids,* 274**,** 470-476.

Rezakazemi, M. & Shirazian, S. 2019. Gas-liquid phase recirculation in bubble column reactors: Development of a hybrid model based on local CFD - Adaptive Neuro-Fuzzy Inference System (ANFIS). *Journal of Non-Equilibrium Thermodynamics,* 44**,** 29-42.

Riahi-Madvar, H., Dehghani, M., Seifi, A., Salwana, E., Shamshirband, S., Mosavi, A. and Chau, K.W., 2019. Comparative analysis of soft computing techniques RBF, MLP, and ANFIS with MLR and MNLR for predicting grade-control scour hole geometry. *Engineering Applications of Computational Fluid Mechanics,* 13(1), pp.529-550.

Rostami, A., Baghban, A., Mohammadi, A. H., Hemmati-Sarapardeh, A. & Habibzadeh, S. J. F. 2019. Rigorous prognostication of permeability of heterogeneous carbonate oil reservoirs: Smart modeling and correlation development. 236**,** 110-123.

Sadrzadeh, M., Mohammadi, T., Ivakpour, J. & Kasiri, N. 2009. Neural network modeling of Pb2+ removal from wastewater using electrodialysis. *Chemical Engineering and Processing: Process Intensification,* 48**,** 1371-1381.

Samavati, V., Emam-Djomeh, Z. & Omid, M. 2013. Prediction of Rheological Properties of Multi-Component Dispersions by Using Artificial Neural Networks. *Journal of Dispersion Science and Technology,* 35**,** 428-434.

Shariati, A. & Peters, C. J. 2005. High-pressure phase equilibria of systems with ionic liquids. *The Journal of Supercritical Fluids,* 34**,** 171-176.

Shamshirband, S.; Baghban, A.; Hadipoor, M.; Mosavi, A. Developing an ANFIS-PSO Based Model to Estimate Mercury Emission in Combustion Flue Gases. Preprints 2019, 2019050124 (doi: 10.20944/preprints201905.0124.v2).

Shi, X. & Che, D. 2009. A combined power cycle utilizing low-temperature waste heat and LNG cold energy. *Energy Conversion and Management,* 50**,** 567-575.

Sipöcz, N., Tobiesen, F. A. & Assadi, M. 2011. The use of Artificial Neural Network models for CO2 capture plants. *Applied Energy,* 88**,** 2368-2376.

Soave, G. 1972. Equilibrium constants from a modified Redlich-Kwong equation of state. *Chemical Engineering Science,* 27**,** 1197-1203.

Suleymani, M. & Bemani, A. 2018. Application of ANFIS-PSO algorithm as a novel method for estimation of higher heating value of biomass. *Energy Sources, Part A: Recovery, Utilization and Environmental Effects,* 40**,** 288-293.
Svozil, D., Kvasnicka, V. & Pospichal, J. í. 1997. Introduction to multi-layer feed-forward neural networks. *Chemometrics and Intelligent Laboratory Systems,* 39**,** 43-62.
Torabi, M., Hashemi, S., Saybani, M. R., Shamshirband, S. & Mosavi, A. 2019. A Hybrid clustering and classification technique for forecasting short-term energy consumption. *Environmental Progress and Sustainable Energy,* 38**,** 66-76.
Veit, D. 2012. 4 - Fuzzy logic and its application to textile technology. *In:* VEIT, D. (ed.) *Simulation in Textile Technology.* Woodhead Publishing.
Wang, C., Cui, G., Luo, X., Xu, Y., Li, H. & Dai, S. 2011. Highly Efficient and Reversible SO2 Capture by Tunable Azole-Based Ionic Liquids through Multiple-Site Chemical Absorption. *Journal of the American Chemical Society,* 133**,** 11916-11919.
Wang, M., He, L. & Infante Ferreira, C. A. 2019. Ammonia absorption in ionic liquids-based mixtures in plate heat exchangers studied by a semi-empirical heat and mass transfer framework. *International Journal of Heat and Mass Transfer***,** 1302-1317.
Welton, T. 1999. Room-Temperature Ionic Liquids. Solvents for Synthesis and Catalysis. *Chemical Reviews,* 99**,** 2071-2084.
Yang, Q. & Dionysiou, D. D. 2004. Photolytic degradation of chlorinated phenols in room temperature ionic liquids. *Journal of Photochemistry and Photobiology A: Chemistry,* 165**,** 229-240.
Yaseen, Z. M., Sulaiman, S. O., Deo, R. C. & Chau, K.-W. J. J. o. h. 2018. An enhanced extreme learning machine model for river flow forecasting: State-of-the-art, practical applications in water resource engineering area and future research direction.
Yilbas, B. S. & Sahin, A. Z. 2014. Thermal characteristics of combined thermoelectric generator and refrigeration cycle. *Energy Conversion and Management,* 83**,** 42-47.
Yokozeki, A. & Shiflett, M. B. 2007a. Ammonia Solubilities in Room-Temperature Ionic Liquids. *Industrial & Engineering Chemistry Research,* 46**,** 1605-1610.
Yokozeki, A. & Shiflett, M. B. 2007b. Vapor–liquid equilibria of ammonia + ionic liquid mixtures. *Applied Energy,* 84**,** 1258-1273.
Yokozeki, A., Shiflett, M. B., Junk, C. P., Grieco, L. M. & Foo, T. 2008. Physical and Chemical Absorptions of Carbon Dioxide in Room-Temperature Ionic Liquids. *The Journal of Physical Chemistry B,* 112**,** 16654-16663.
Yokozeki, A., Shiflett, M. B. J. I. & research, e. c. 2007. Ammonia solubilities in room-temperature ionic liquids. 46**,** 1605-1610.
Zendehboudi, S., Ahmadi, M. A., James, L. & Chatzis, I. 2012. Prediction of Condensate-to-Gas Ratio for Retrograde Gas Condensate Reservoirs Using Artificial Neural Network with Particle Swarm Optimization. *Energy & Fuels,* 26**,** 3432-3447.
Zeng, S., Wang, J., Li, P., Dong, H., Wang, H., Zhang, X. & Zhang, X. 2019. Efficient adsorption of ammonia by incorporation of metal ionic liquids into silica gels as mesoporous composites. *Chemical Engineering Journal***,** 81-88.

**Table 1** Ammonia solubility in different ranges of temperatures, pressures and ILs.

| No. | Ionic liquid | Temperature range (K) | Pressure range (MPa) | $NH_3$ solubility range (mole fraction) | No. of data points | References |
|---|---|---|---|---|---|---|
| 1 | [bmim][PF6] | 283.4-355.8 | 0.138-2.7 | 0.239-0.862 | 29 | (Peng and Robinson, 1976) |
| 2 | [bmim][BF4] | 282.2-355.1 | 0.07-2.57 | 0.0608-0.883 | 61 | (Soave, 1972, Marquardt, 1963) |
| 3 | [emim][Tf2N] | 283.3-347.6 | 0.114-2.86 | 0.045-0.948 | 30 | (Yokozeki and Shiflett, 2007a) |
| 4 | [hmim][Cl] | 283.1-347.9 | 0.044-2.49 | 0.06-0.837 | 30 | (Yokozeki and Shiflett, 2007a) |
| 5 | [C2MIM][BF4] | 293.15-333.15 | 0.11-0.63 | 0.0838-0.6921 | 25 | (Yokozeki and Shiflett, 2007b) |
| 6 | [C6MIM][BF4] | 293.15-333.15 | 0.13-0.71 | 0.128-0.7531 | 25 | (Yokozeki and Shiflett, 2007a) |
| 7 | [C8MIM][BF4] | 293.15-333.15 | 0.1-0.61 | 0.1321-0.8081 | 25 | (Yokozeki and Shiflett, 2007a) |
| 8 | [emim][Ac] | 282.5-348.5 | 0.321-2.891 | 0.473-0.877 | 30 | (Yokozeki and Shiflett, 2007b) |
| 9 | [emim][EtOSO3] | 282.7.15-372.3 | 0.287-4.777 | 0.424-0.875 | 29 | (Yokozeki and Shiflett, 2007b) |
| 10 | [emim][SCN] | 283.2-372.8 | 0.244-5.007 | 0.34-0.876 | 36 | (Yokozeki and Shiflett, 2007b) |
| 11 | [DMEA][Ac] | 283.2-372.8 | 0.136-4.249 | 0.454-0.865 | 32 | (Chen et al., 2013) |

**Table 2** Specifications of ionic liquids.

| No. | Ionic liquid | Mw (Kg/Kmole) | Tc (K) | $P_c$ (MPa) |
|---|---|---|---|---|
| 1 | [bmim][PF6] | 284.18 | 860.5 | 2.645 |
| 2 | [bmim][BF4] | 226.02 | 894.9 | 3.019 |
| 3 | [emim][Tf2N] | 391.31 | 808.8 | 2.028 |
| 4 | [hmim][Cl] | 202.73 | 951.3 | 3.061 |
| 5 | [C2MIM][BF4] | 198 | 585.3 | 2.36 |
| 6 | [C6MIM][BF4] | 254 | 679.1 | 1.79 |
| 7 | [C8MIM][BF4] | 282.1 | 726.1 | 1.6 |
| 8 | [emim][Ac] | 170.11 | 871.3 | 3.595 |
| 9 | [emim][EtOSO3] | 236.29 | 945.4 | 3.122 |
| 10 | [emim][SCN] | 169.25 | 1001.1 | 4.102 |
| 11 | [DMEA][Ac] | 149.19 | 727.9 | 3.397 |

**Table 3** Main parameters of PSO-ANFIS model.

| Type | Value/comment |
| --- | --- |
| **Membership function** | Gaussmf |
| **Maximum iteration** | 2500 |
| **Minimum amount of improvement** | 1e-5 |
| **Number of data used for training** | 264 |
| **Number of data used for testing** | 88 |
| **Optimization method** | PSO |
| **Train Epoch** | 200 |
| **Number of cluster** | 10 |
| **Number of FIS parameter** | 120 |

**Table 4** main features of MLP-ANN approach.

| Type | Value/comment |
|---|---|
| **Input layer** | 5 |
| **Hidden layer** | 10 |
| **Output layer** | 1 |
| **Hidden layer activation function** | Logsig |
| **Output layer activation function** | Purelin |
| **Optimization method** | BP |
| **Number of data used for training** | 264 |
| **Number of data used for testing** | 88 |
| **Number of max iterations** | 2500 |

**Table 5** Weighting values and bias amounts of the proposed MLP-ANN network.

| Neuron | Interior Layers | | | | | | Outer Layer | |
|---|---|---|---|---|---|---|---|---|
| | Weights | | | | | Bias | Weights | Bias |
| | T | P | $T_c$ | $P_c$ | Mw | B1 | x | B2 |
| 1 | 18.86894 | -14.9647 | 30.48689 | -91.0551 | -5.92461 | 45.65722 | 0.242268 | -94.7474617 |
| 2 | 9.900296 | 9.552165 | -6.00342 | 14.03248 | -7.28674 | -10.2618 | -0.74979 | |
| 3 | -10.8064 | 30.0597 | -2.12182 | 5.167274 | -3.43343 | 14.97859 | -65.0436 | |
| 4 | 1.727799 | 8.276695 | -0.39079 | -0.41015 | -0.38428 | 1.903079 | 161.453 | |
| 5 | -1.10441 | 1.788451 | -41.9287 | -18.4621 | 32.63778 | 0.029692 | 2.555817 | |
| 6 | -14.1359 | 3.808112 | -8.85667 | -12.3691 | -6.41565 | 28.29349 | -0.7418 | |
| 7 | 3.094511 | 10.39778 | 1.278268 | -2.08559 | -0.03795 | -3.75952 | -0.61798 | |
| 8 | -17.7873 | 9.299117 | 48.50644 | -85.318 | 49.71805 | 9.706551 | -0.33592 | |
| 9 | -7.22462 | 0.980609 | 0.102641 | -1.11196 | -0.17822 | 5.278534 | 5.346496 | |
| 10 | -1.89945 | -9.83265 | 0.330473 | 0.84824 | 0.359386 | -0.90802 | 48.92541 | |

**Table 6** statistical analysis of performances.

| Network | PSO-ANFIS | | MLP-ANN | | EOS | |
|---|---|---|---|---|---|---|
| **Analysis** | Training | Testing | Training | Testing | PR | SRK |
| **MSE** | 0.002007 | 0.004577 | 0.000223 | 0.000453 | 0.010156 | 0.102345 |
| **$R^2$** | 0.9618 | 0.9125 | 0.9967 | 0.992 | 0.8681 | 0.4225 |
| **RMSE** | 0.044797 | 0.067657 | 0.014938 | 0.021295 | 0.100778 | 0.319915 |
| **AAE %** | 1.629359 | 0.742338 | 1.23697 | 0.001013 | 5.63 | 22.49 |
| **STD** | 0.044856 | 0.067971 | 0.014949 | 0.02137 | 0.092694 | 0.196344 |

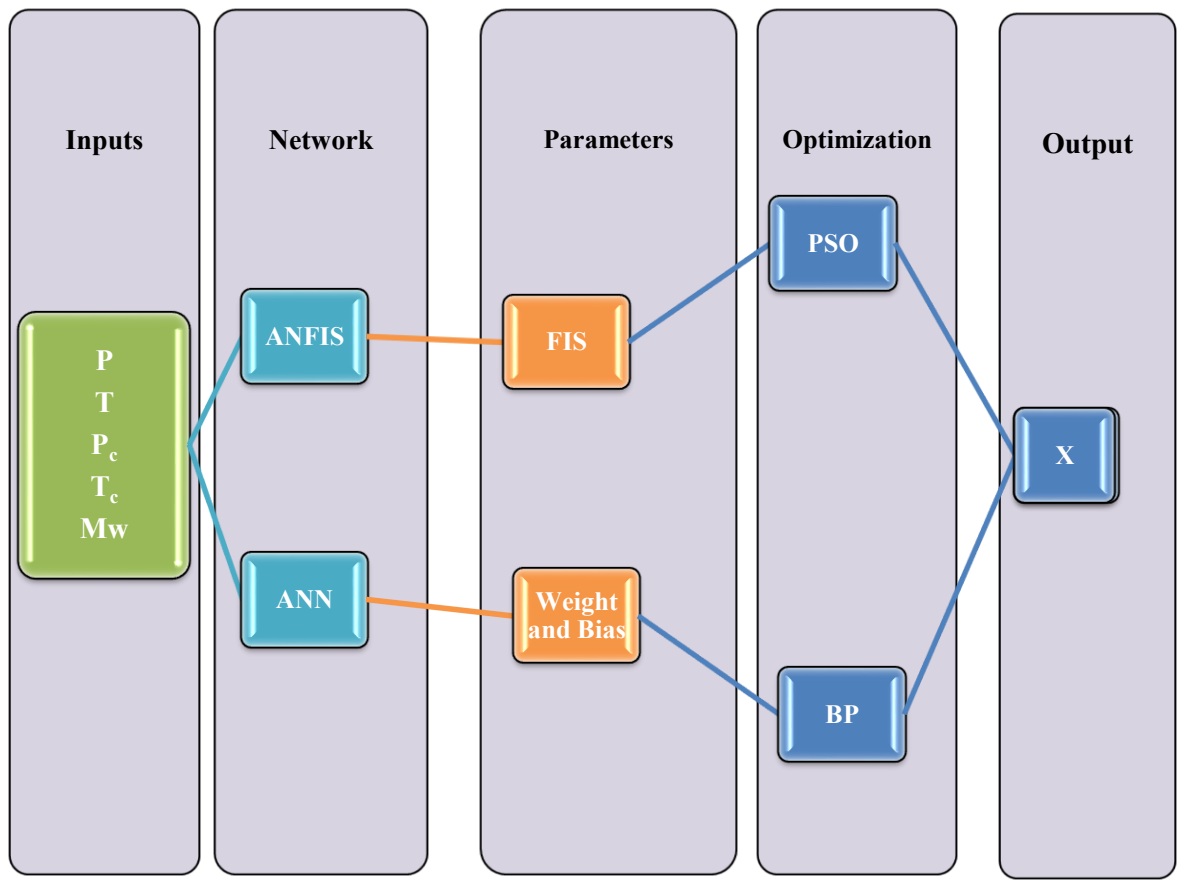

**Figure 1:** Diagram of proposed ANN and ANFIS for NH$_3$ solubility in ILs

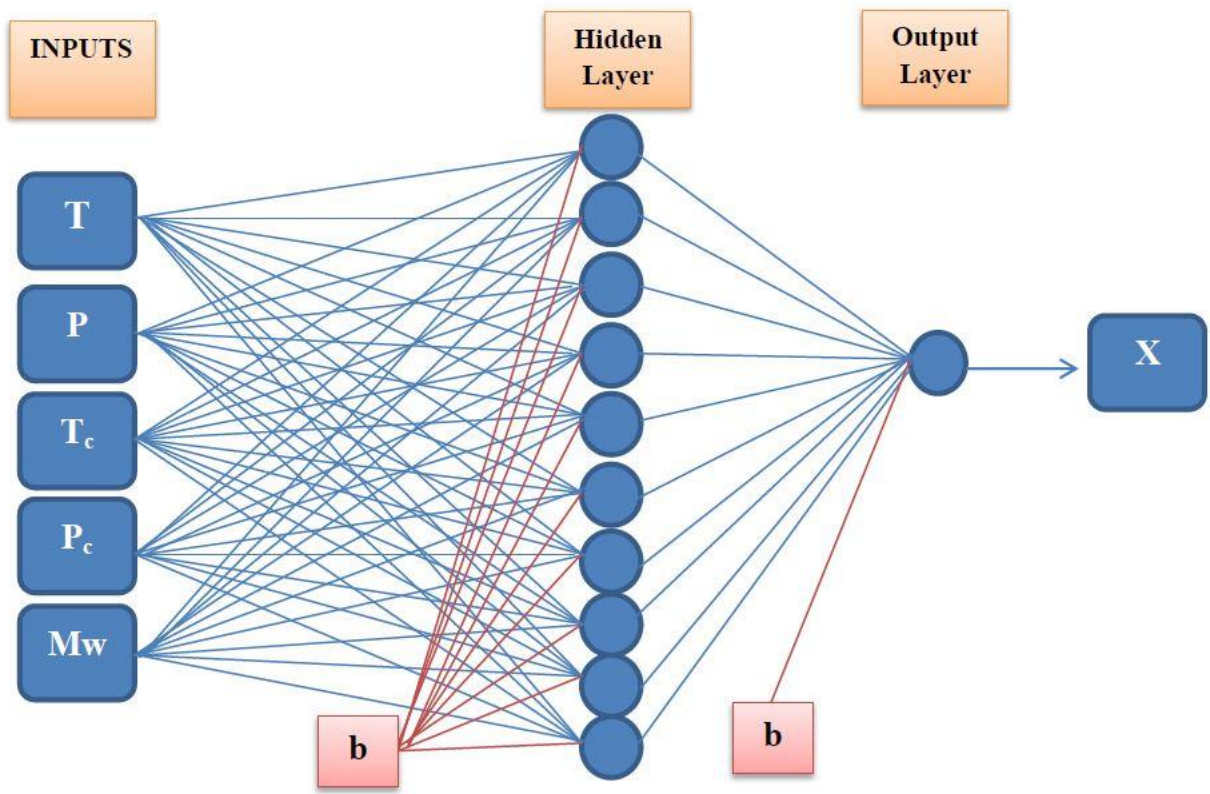

**Figure 2:** schematic of the MLP-ANN model which is used in present study.

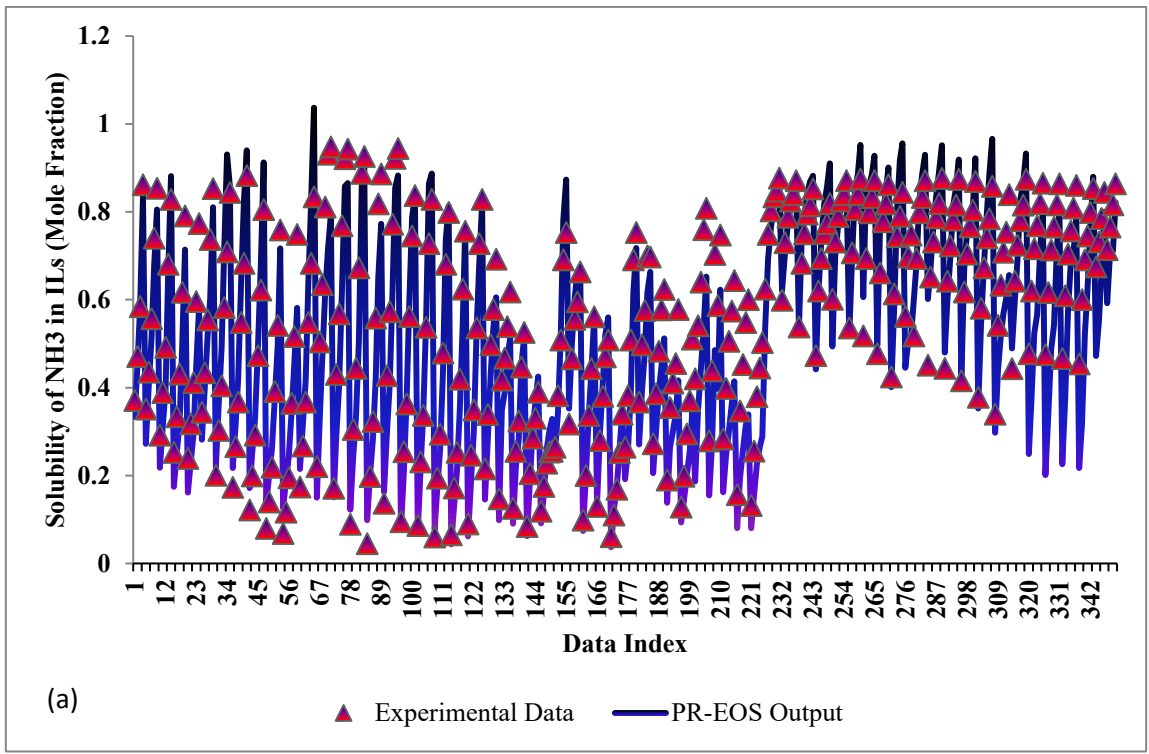

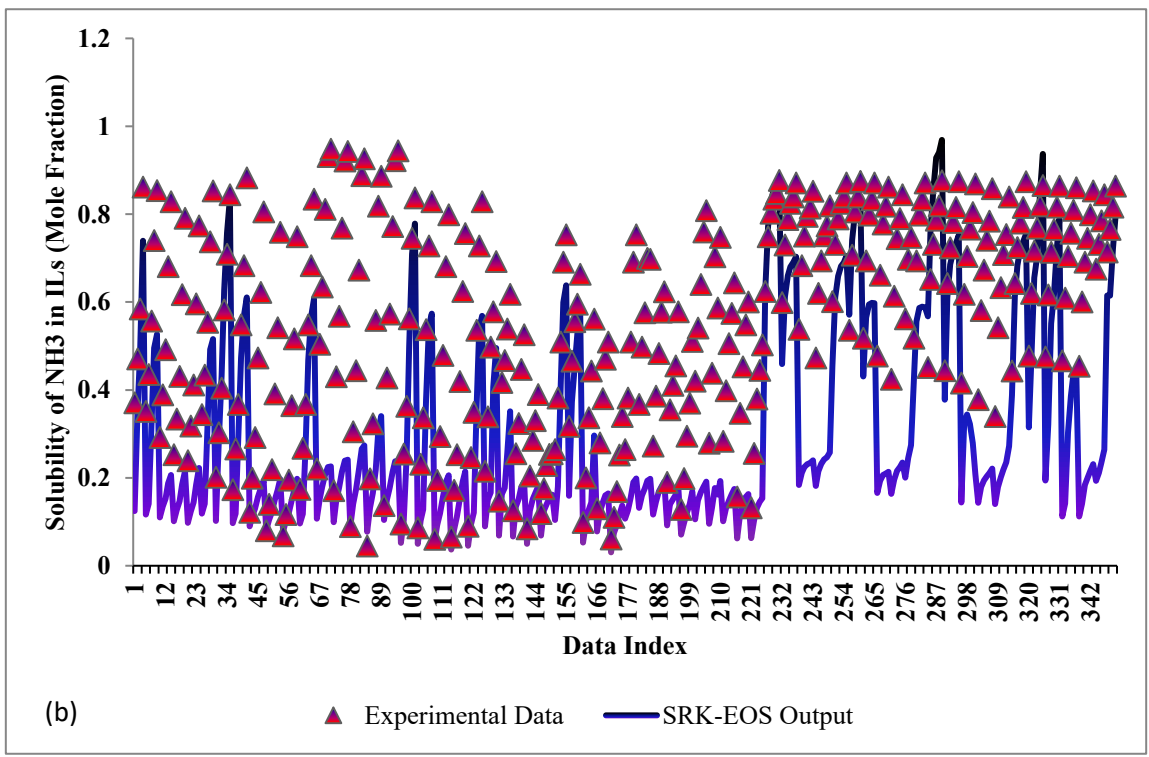

**Figure 3:** Real and predicted amounts of dissolved $NH_3$ in ILs: (a) PR-EOS and (b) SRK-EOS.

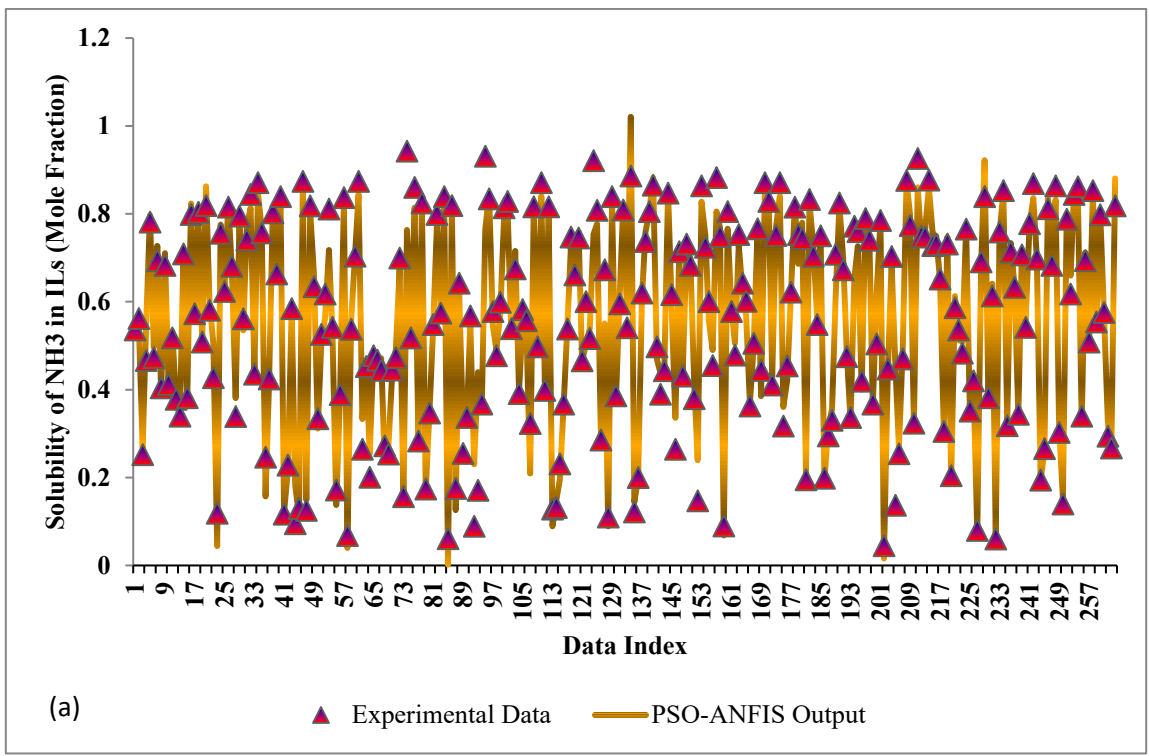
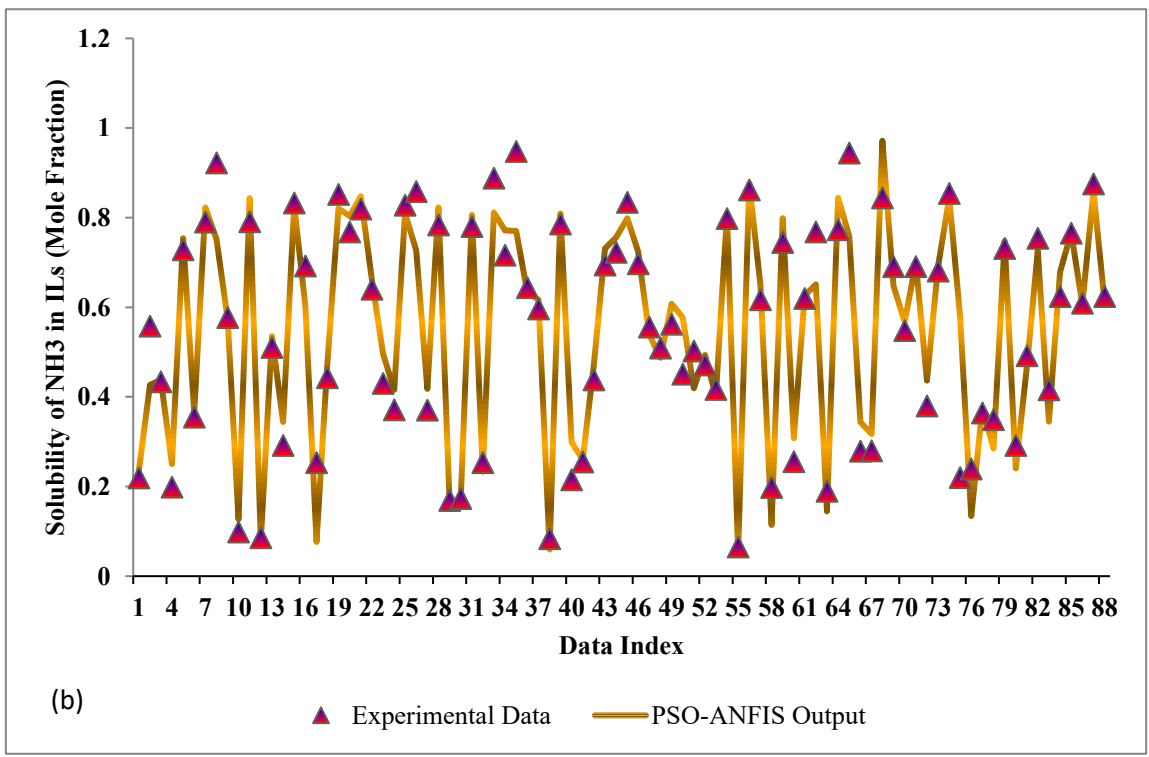

**Figure 4:** Dissolved mole fractions of NH$_3$ in ILs, estimated by PSO-ANFIS at: (a) Training phase (b) Testing stage.

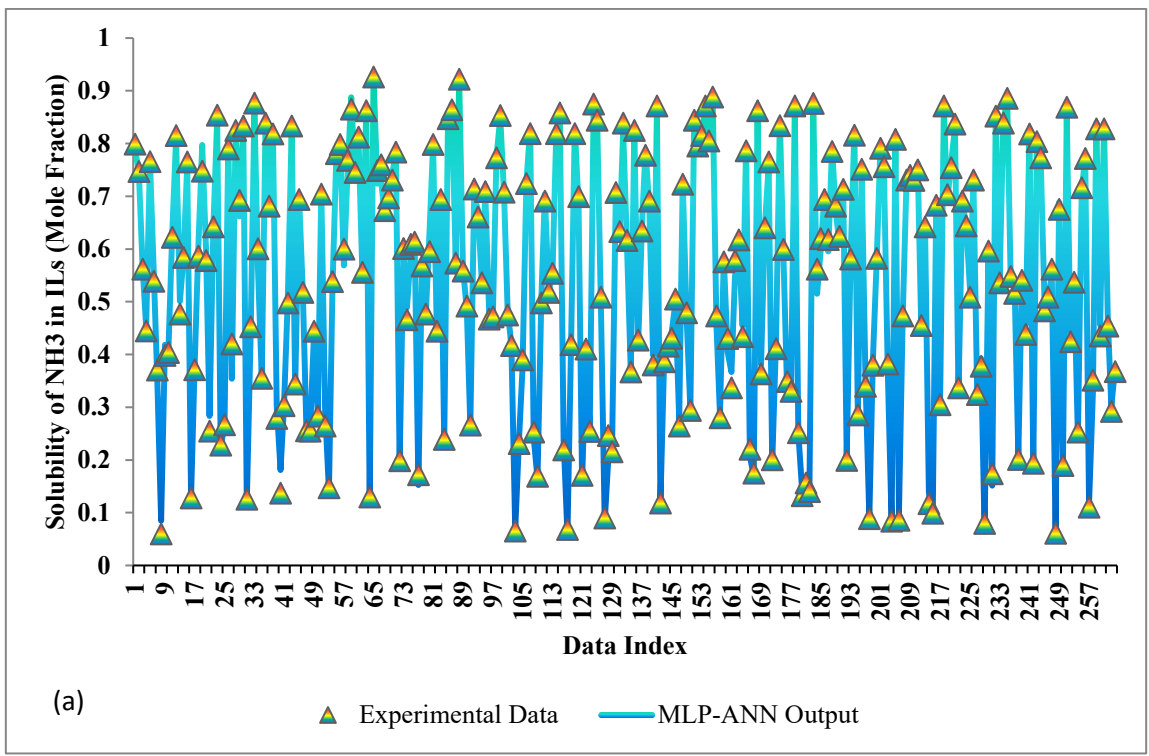

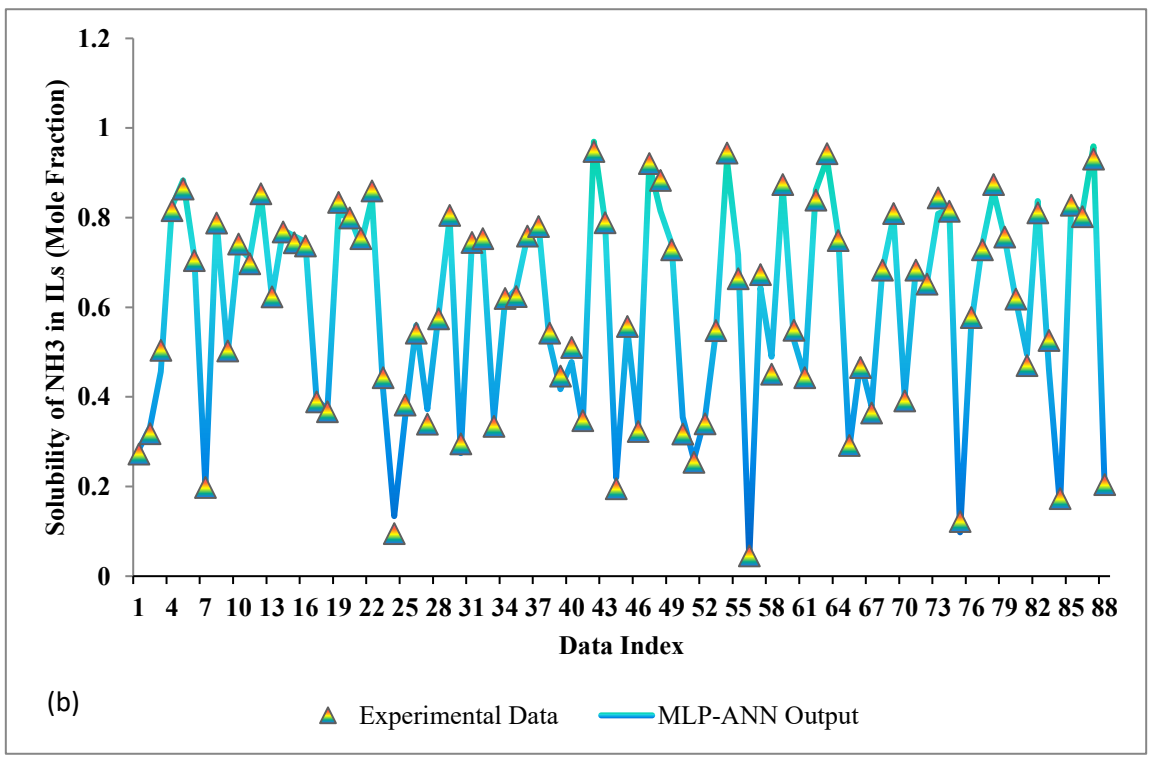

**Figure 5:** Dissolved mole fractions of NH$_3$ in ILs, estimated by MLP-ANN at: (a) Training phase (b) Testing stage.

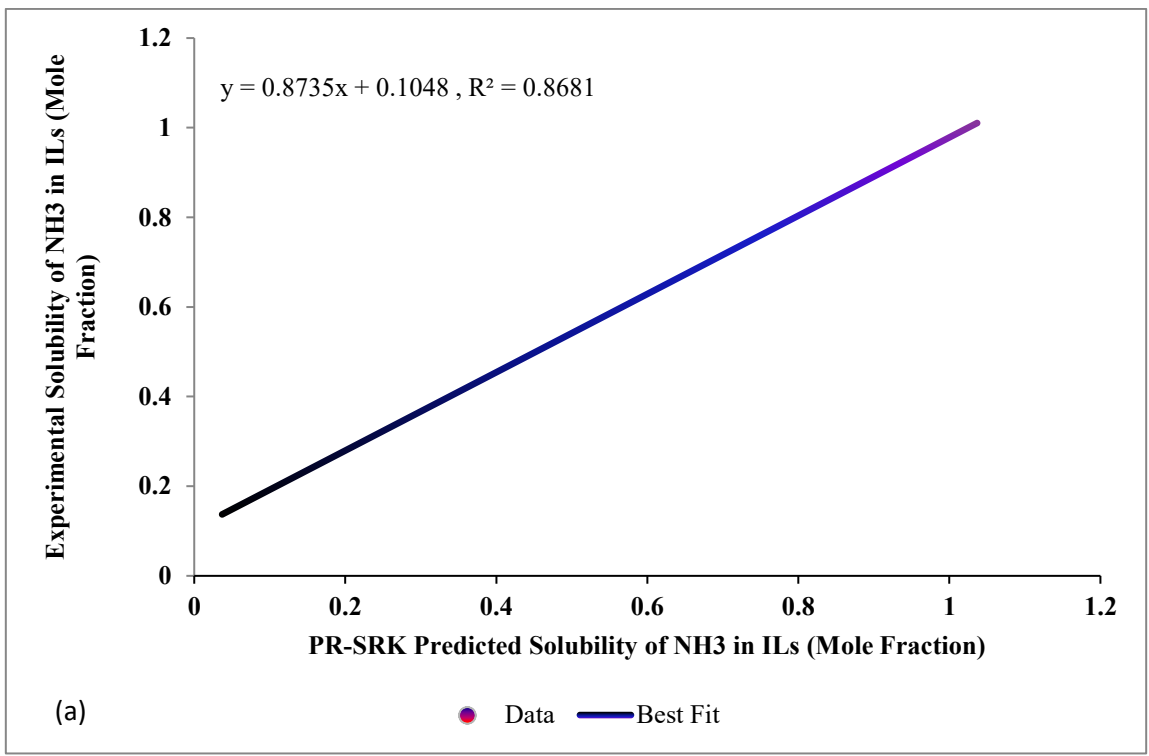

(a)

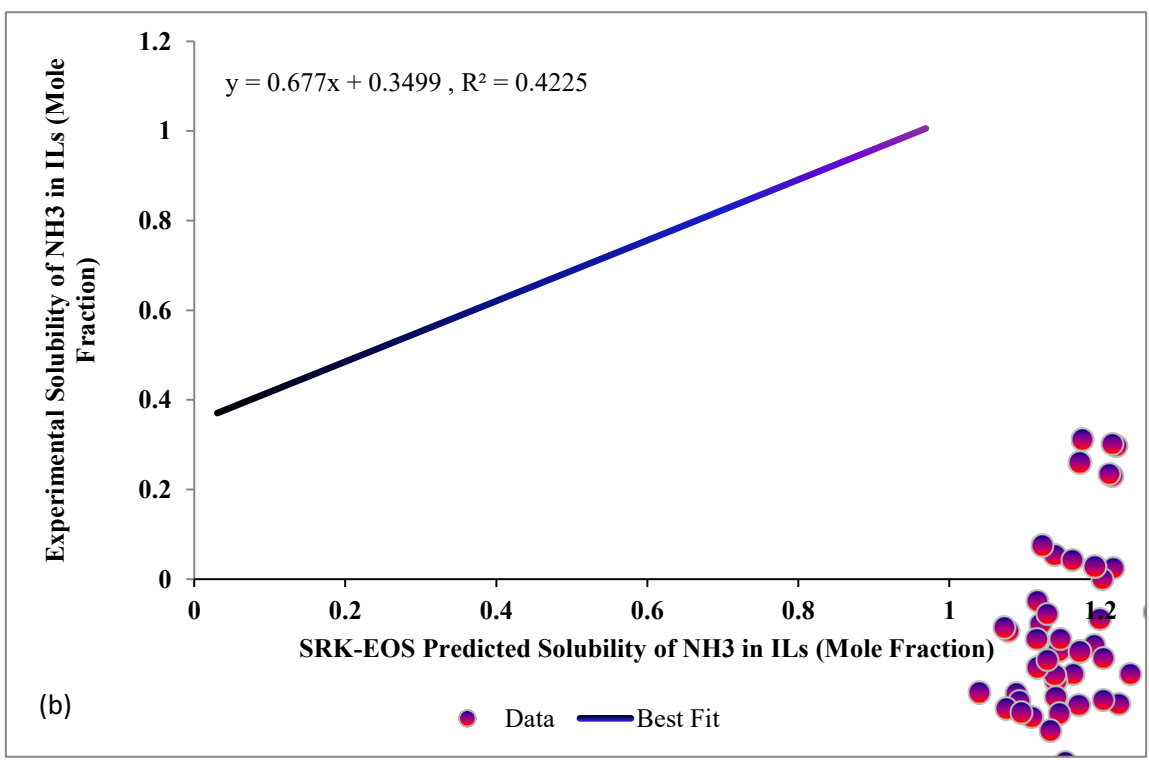

(b)

**Figure 6:** Regression plots for the solubility of NH$_3$, estimated values by EOS versus experimental data: a) PR-EOS b) SRK-EOS.

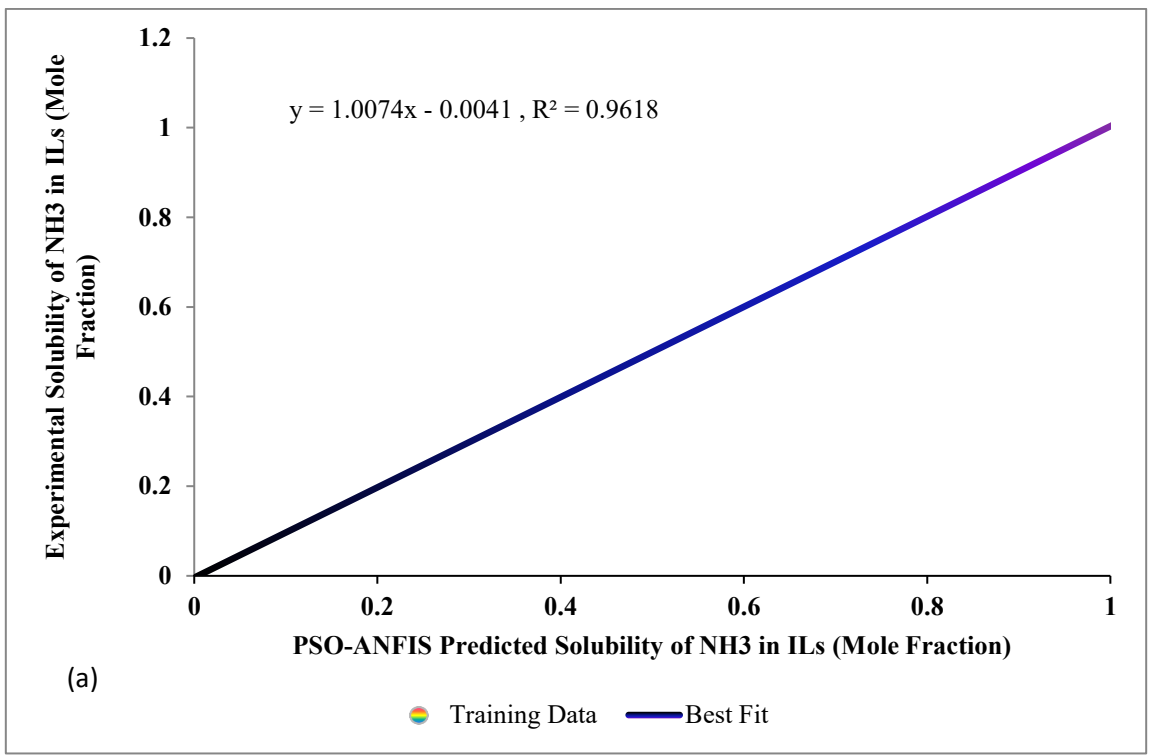

(a)

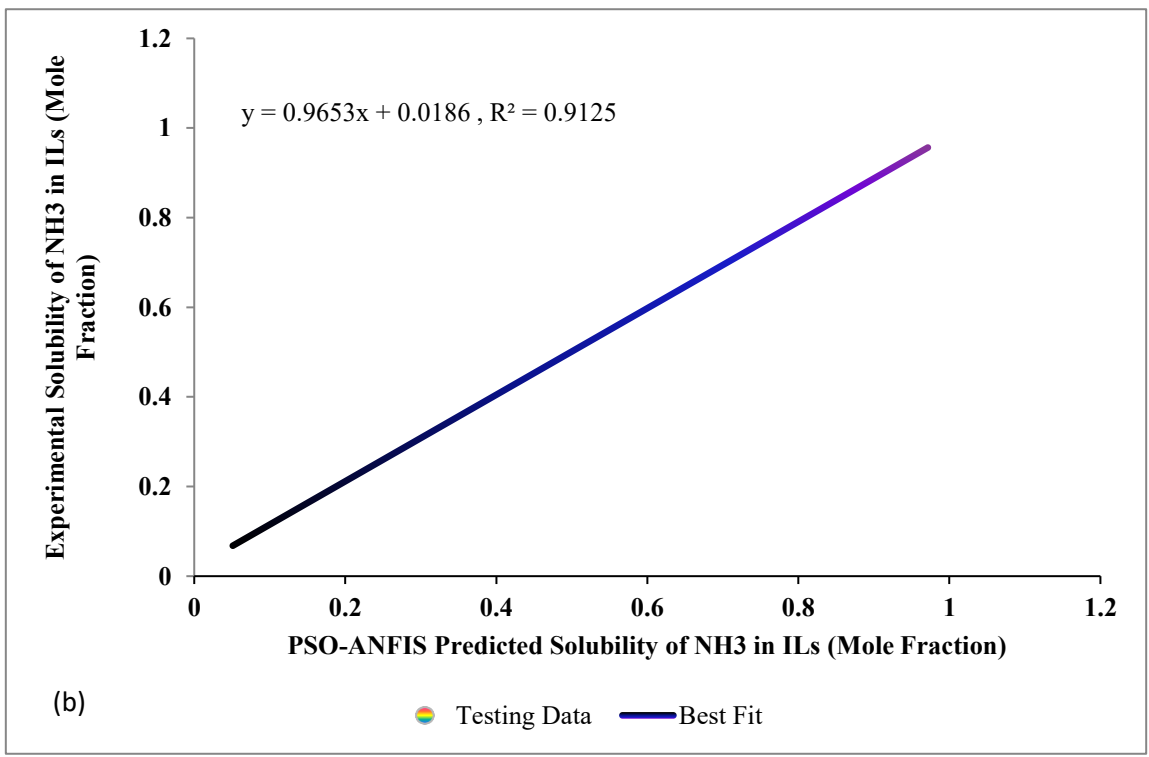

(b)

**Figure 7:** Regression plots of PSO-ANFIS model at: (a) Training and (b) Testing stages.

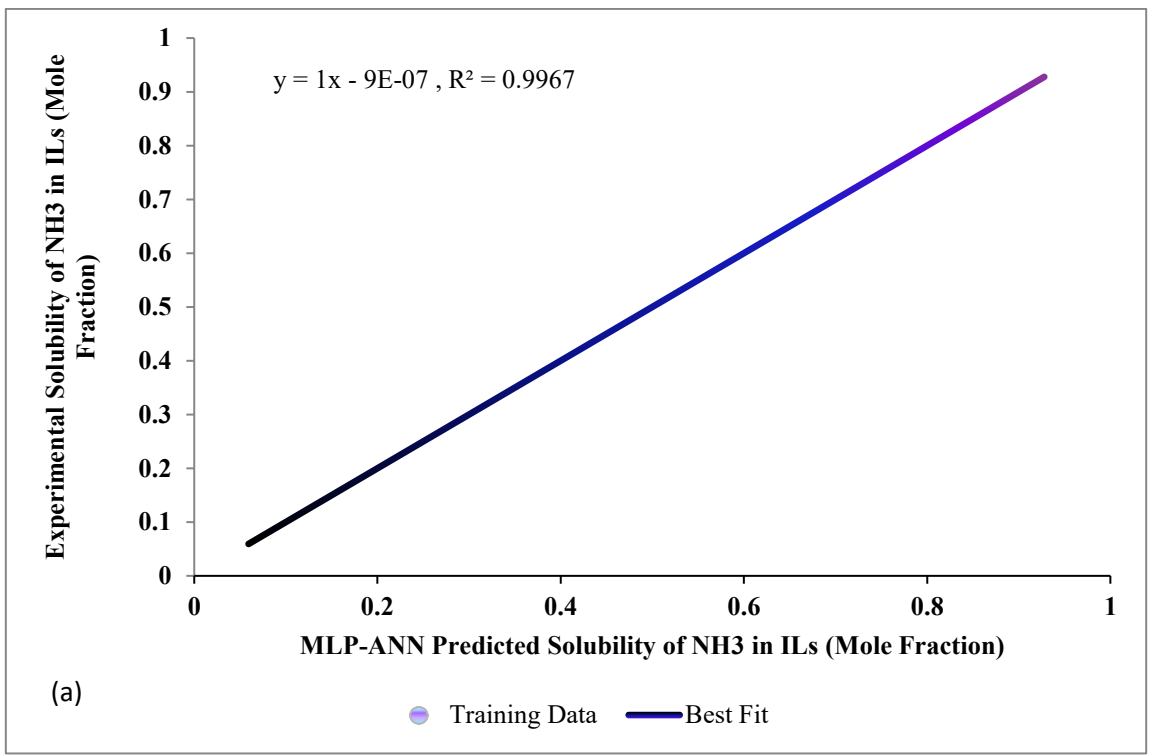

(a)

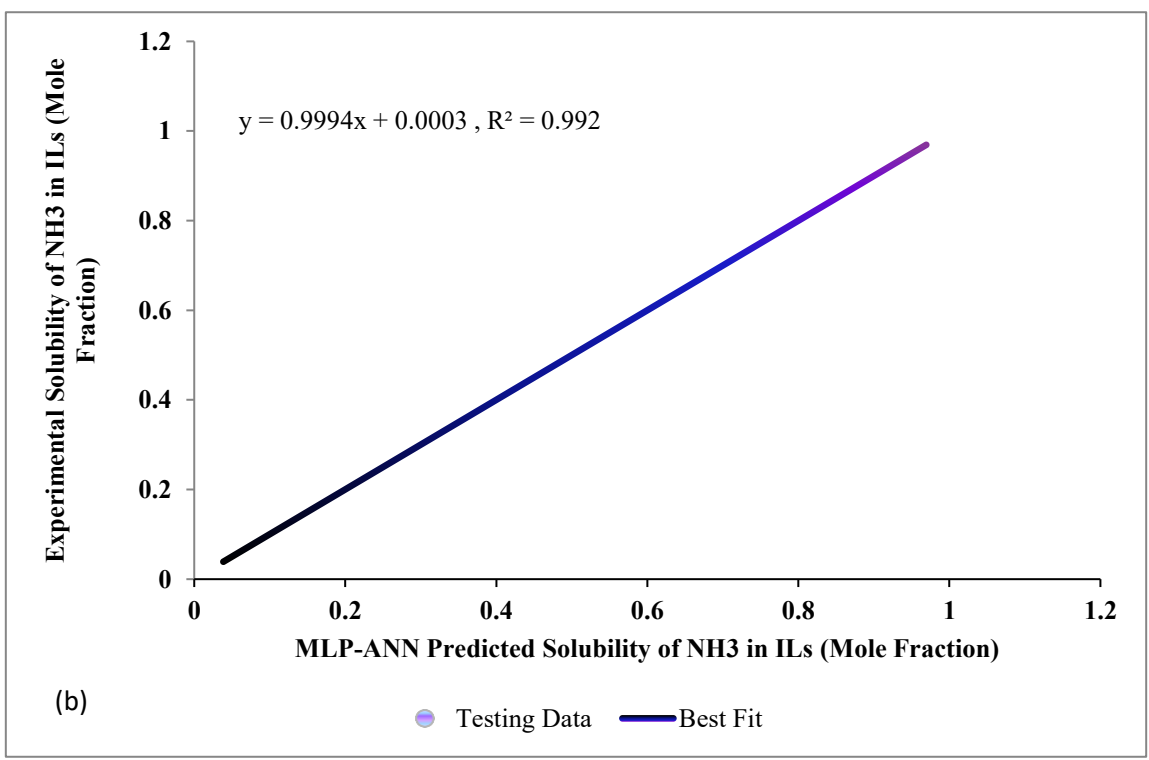

(b)

**Figure 8:** Regression plots of MLP-ANN model at: (a) Training and (b) Testing stages.

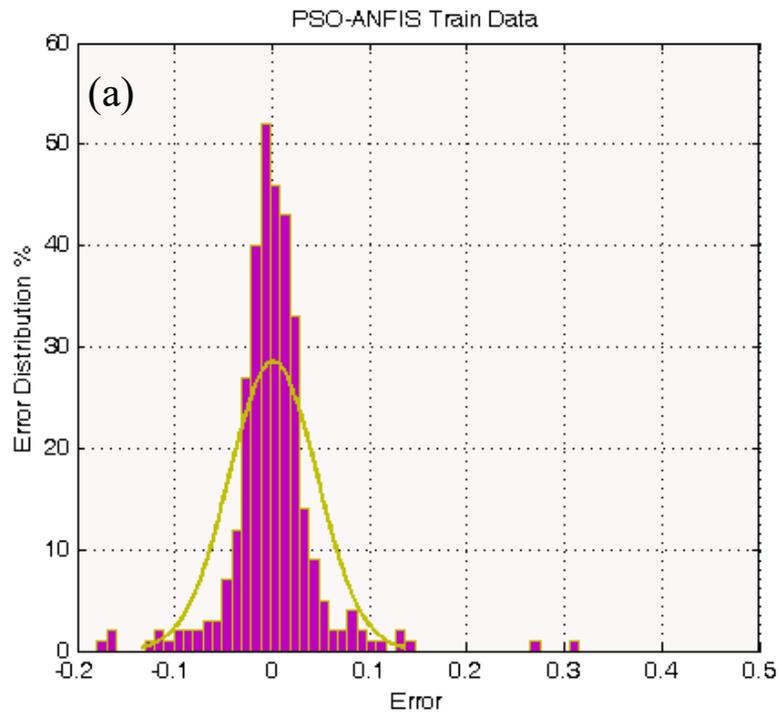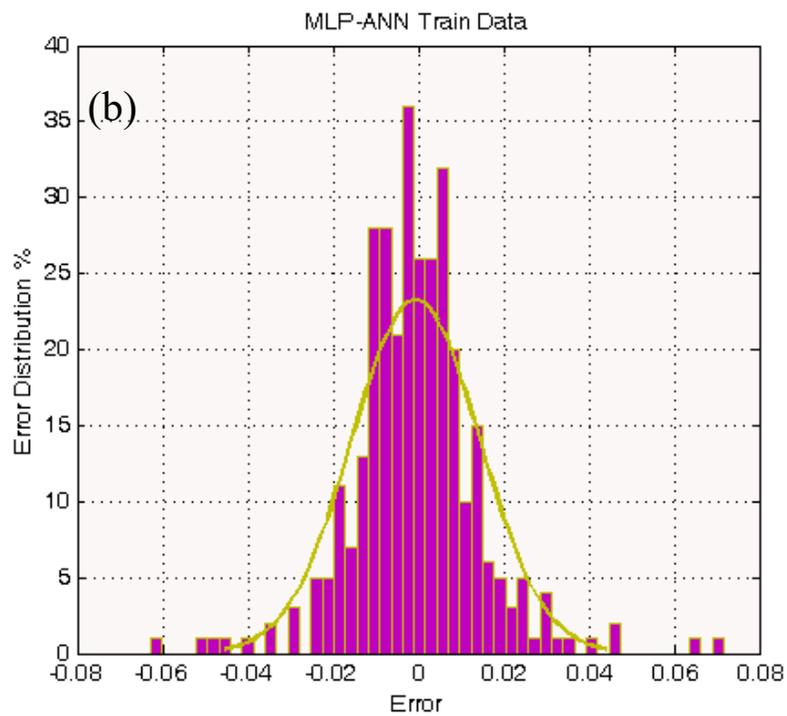

**Figure 9:** Histogram of errors during the estimation of the Ammonia solubility: (a) PSO-ANFIS and (b) MLP-ANN

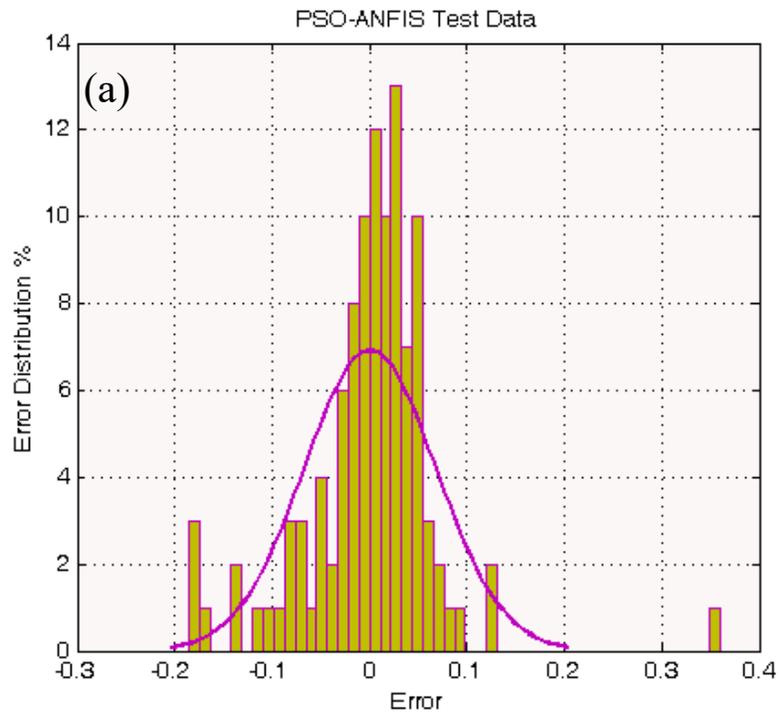

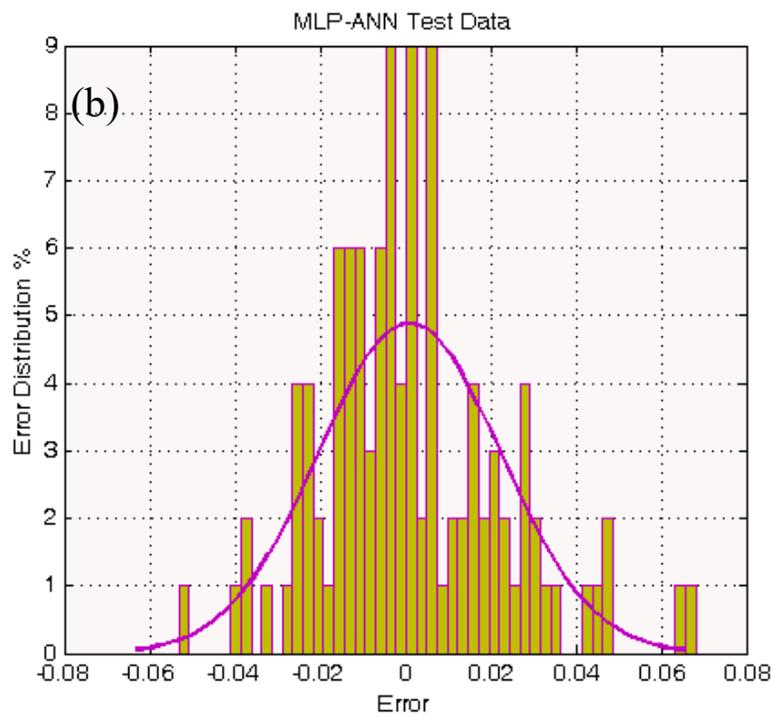

**Figure 10:** Histogram of errors during the estimation of the Ammonia solubility: (a) PSO-ANFIS and (b) MLP-ANN.

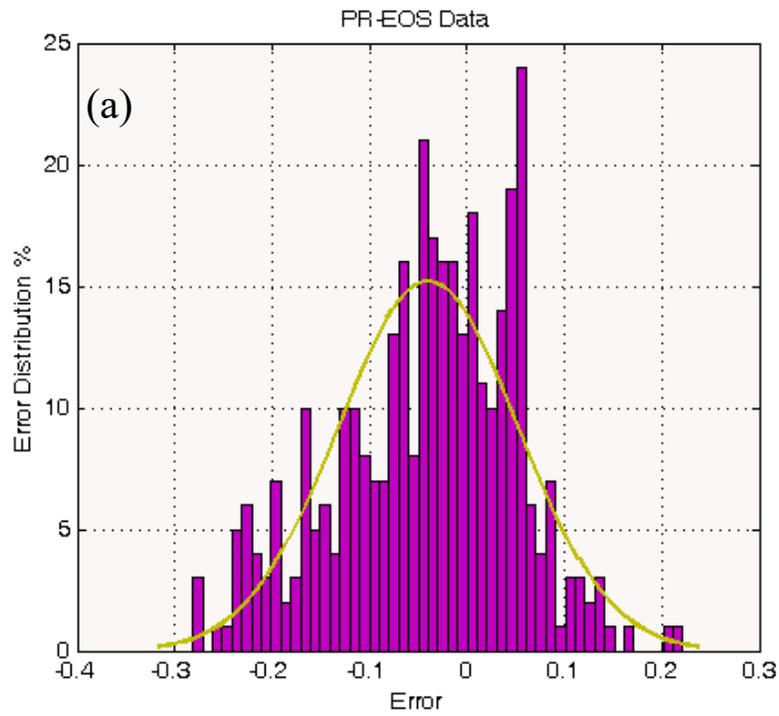

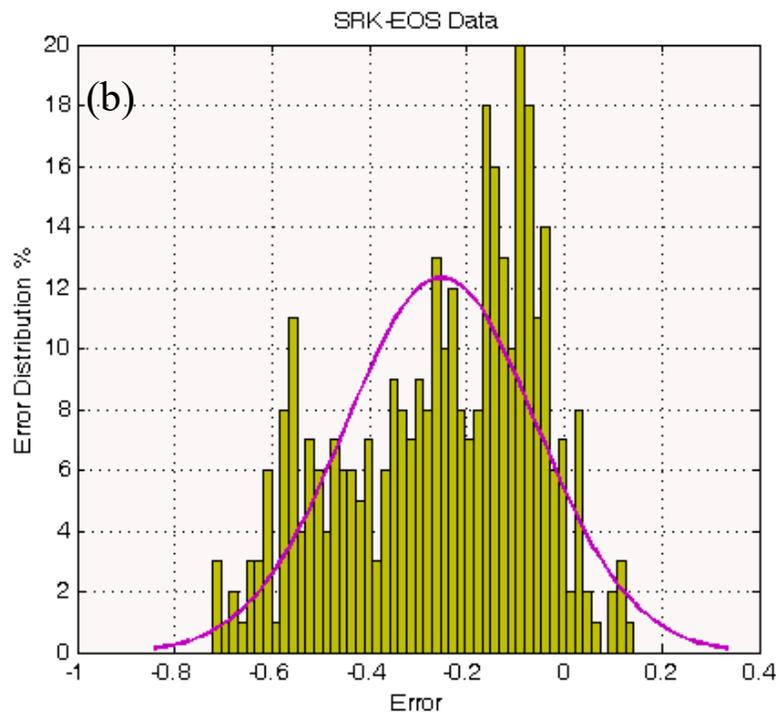

**Figure 11:** Histogram of errors during the estimation of the Ammonia solubility: (a) PR-EOS and (b) SRK-EOS.

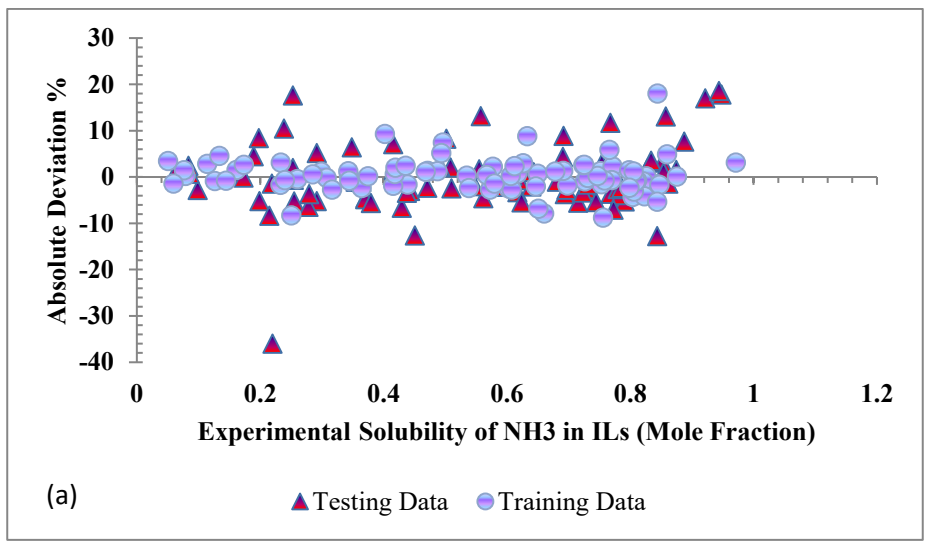

(a)

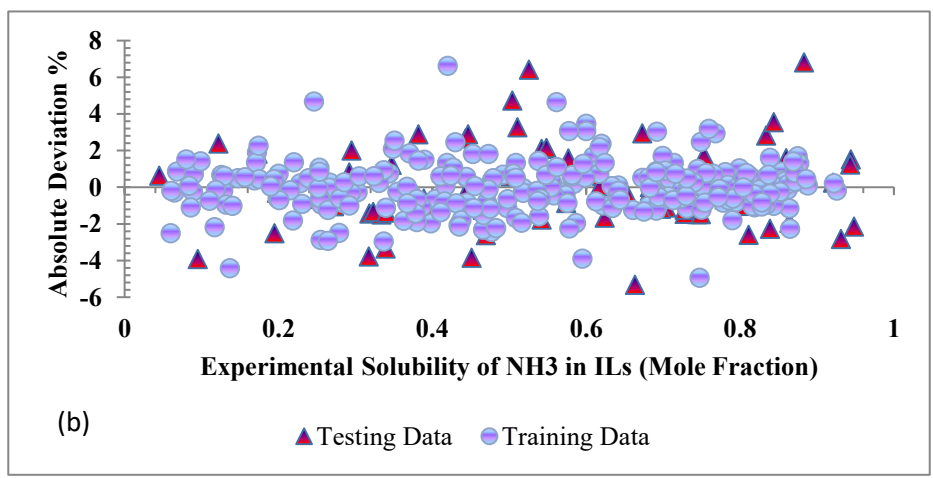

(b)

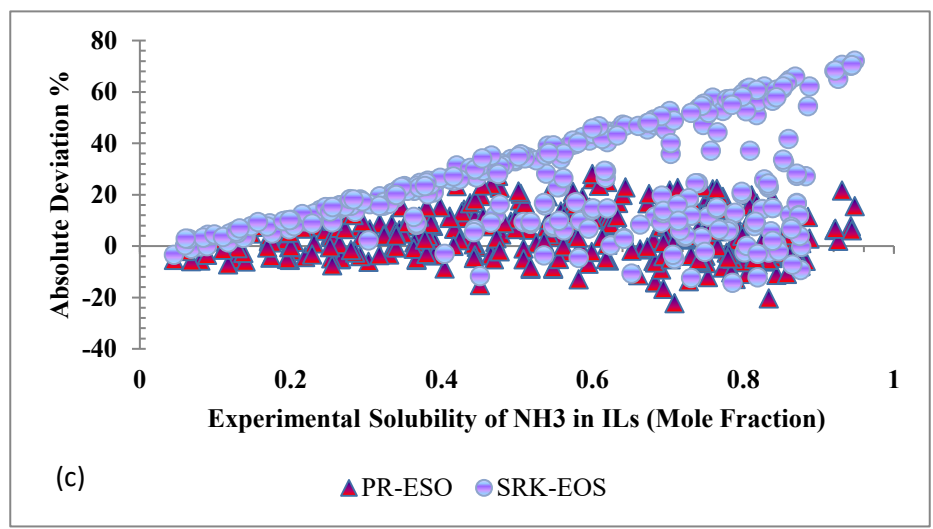

(c)

**Figure 12:** Absolute Deviation (%) of estimated $NH_3$ solubility using: (a) PSO-ANFIS, (b) MLP-ANN and (c) EOS (PR-EOS and SRK-EOS) versus experimental ones.